\documentclass[12pt]{article}

\usepackage[T1]{fontenc}

\usepackage{amsmath}
\usepackage{amssymb}
\usepackage{slashed}
\usepackage{tikz}
\usepackage[numbers,sort&compress]{natbib}
\usepackage{hyperref}
\usepackage{cleveref}
\crefname{equation}{Eq.}{Eqs.}
\crefname{figure}{Fig.}{Figs.}
\crefname{table}{Table}{Tables}
\crefname{section}{Section}{Sections}

\usepackage[letterpaper,margin=1in,bottom=1in]{geometry}
\usepackage{float} % allows forcing plot and table locations with [H]
\usepackage{parskip} % skip spaces between paragraphs
\usepackage{tabulary} % table environment with auto word wrapping
\usepackage{color} % color for tracking edits
\usepackage{soul} % strikeout for tracking edits
\usepackage{subfig}
\usepackage{graphicx}
\usepackage[section]{placeins} % keep figures inside their sections
\usepackage{multirow}

\def\lhc2{LHC~Run~II}

\def\beq{\begin{equation}}
\def\be{\begin{equation}}
\def\beqn{\begin{eqnarray}}
\def\ee{\end{equation}}
\def\eeq{\end{equation}}
\def\eeqn{\end{eqnarray}}

\begin{document}

\author{Amin Aboubrahim$^a$\footnote{Email: aabouibr@uni-muenster.de}, ~Michael Klasen$^a$\footnote{Email: michael.klasen@uni-muenster.de}, ~and Pran Nath$^b$\footnote{Email: p.nath@northeastern.edu}\\~\\
$^{a}$\textit{\normalsize Institut f\"ur Theoretische Physik, Westf\"alische Wilhelms-Universit\"at M\"unster,} \\ 
\textit{\normalsize Wilhelm-Klemm-Stra{\ss}e 9, 48149 M\"unster, Germany}\\
$^{b}$\textit{\normalsize Department of Physics, Northeastern University, Boston, MA 02115-5000, USA}} 
\title{\vspace{-2.0cm}\begin{flushright}
{\small MS-TP-20-42}
\end{flushright}
\vspace{1cm}
Xenon-1T excess as a possible signal of a sub-GeV  hidden sector dark matter} 

\date{}
\maketitle

\begin{abstract}
We present a particle physics model to explain the observed  enhancement in the Xenon-1T data
at an electron recoil energy of 2.5 keV. The model is based on a $U(1)$ extension of the
Standard Model where the dark sector consists of two essentially mass degenerate
 Dirac fermions  in the sub-GeV region
 with a small mass splitting interacting with a dark photon. 
 The dark photon is unstable and decays before the big bang nucleosynthesis,  
  which leads to the dark matter constituted of two essentially mass degenerate Dirac fermions.
 The Xenon-1T excess is computed via the inelastic exothermic scattering of the heavier dark
 fermion from a bound electron in xenon to the lighter dark fermion producing the
 observed excess events in the recoil electron energy. The model can be tested with further 
 data from Xenon-1T and in future experiments such as SuperCDMS.

\end{abstract}

\section{Introduction}

Recently the Xenon-1T experiment~\cite{Aprile:2020tmw}
has analyzed events in the low-energy region of 1--30\,keV of electron recoil energy
with an exposure of 0.65 tonne-years, while claiming a low background rate of \mbox{$76 \pm 2_{\,\mathrm{stat}}$\,events/(tonne-year-keV)}. The experiment observed an excess  of recoil electrons over the background in the 2$-$3 keV range. 
 The collaboration analyzed the axion couplings to electrons, photons and nucleons,  and 
the neutrino magnetic moment as possible sources for the signal. However, these models
appear to be in strong tension with stellar constraints~\cite{Viaux:2013lha,Bertolami:2014wua,Battich:2016htm,Giannotti:2017hny}. Another possible source of this excess
is traces of tritium in xenon of size  $(6.2 \pm 2.0) \times 10^{-25}$\,{mol/mol}. The experiment
currently can neither  confirm nor exclude such a possibility. 
 Since the publication of the Xenon-1T results, a variety of models have been proposed which
include light  sterile neutrinos~\cite{Shakeri:2020wvk,Khruschov:2020cnf,Arcadi:2020zni,Khan:2020vaf}, 
a goldstino~\cite{Cao:2020oxq}, an inflaton~\cite{Takahashi:2020uio},
string-motivated models~\cite{Karozas:2020pun,Anchordoqui:2020tlp}, boosted dark matter~\cite{Jho:2020sku,Fornal:2020npv,DelleRose:2020pbh}, 
 and a variety of other models~\cite{Millea:2020xxp,Arias-Aragon:2020qtn,Long:2020uyf,Athron:2020maw,Li:2020naa,Cacciapaglia:2020kbf,Gao:2020wer,Chiang:2020hgb,Choi:2020kch,Okada:2020evk,Baek:2020owl,Gao:2020wfr,Lindner:2020kko,Bramante:2020zos,AristizabalSierra:2020edu,Harigaya:2020ckz,Bell:2020bes,Du:2020ybt,Chakraborty:2020vec,Borah:2020jzi,Bally:2020yid,Kim:2020aua,Farzan:2020dds,Choudhury:2020xui,Babu:2020ivd}. 

In this work we discuss the possibility that the observed effect is a signal from dark matter in the 
 hidden sector. While there are models in the literature where the hidden sector is used for the
 explanation of the Xenon-1T excess, our analysis differs from them in several ways. In our analysis
 we use the Stueckelberg extension of 
 the Standard Model where the Stueckelberg sector consists of an additional $U(1)$ gauge boson interacting via kinetic mixing with the visible sector.
 In many previous works a single Dirac fermion is used which is then
 split into two Majoranas which are given different masses~\cite{Harigaya:2020ckz,Bramante:2020zos}.
 In our analysis we consider two Dirac fermions in the hidden sector carrying $U(1)$ charges with a small mass splitting and interacting
 with the dark photon generated by the Stueckelberg
 mechanism. In the analysis, both the freeze-in and freeze-out mechanisms operate to 
 generate the desired relic density. The analysis given here satisfies all the relevant constraints
on kinetic mixing between the hidden sector and the visible sector
and on the dark photon mass  from the 
 CRESST 2019 DM-nucleon scattering cross section,  from  the neutrino experiment CHARM~\cite{Bergsma:1985is} whose results are reinterpreted as limits on the dark photon~\cite{Gninenko:2012eq}
 as well as from the Planck relic density experiment. 
 The outline of the rest of the paper is as follows: In  section~\ref{sec:model} 
 we discuss the  model to explain the Xenon-1T result.  In section~\ref{sec:dm} an analysis of the
 dark matter relic density is given.
 A discussion of  the inelastic  dark matter-electron scattering is given in section~\ref{sec:e-dm}.
 Event detection rates in the Xenon-1T detector are discussed in section~\ref{sec:detection}.
 Constraints on the model and a fit to the Xenon-1T data is given in section~\ref{sec:fit}.
 Our conclusions are given in section~\ref{sec:conclu}. Several details of the calculation 
 are given in the Appendix.

\section{Stueckelberg extension with hidden sector dark fermions }\label{sec:model}

We extend the Standard Model (SM) gauge group by an extra $U(1)_X$ under which the SM is neutral. The extra gauge field $C^{\mu}$ mixes with the SM $U(1)_Y$ hypercharge $B^{\mu}$ 
via kinetic mixing~\cite{Holdom:1985ag}.
Further, we use the Stueckelberg mechanism~\cite{Kors:2004dx}
 to generate mass for the gauge boson of the hidden sector. The total Lagrangian is then given by
\begin{equation}
\mathcal{L}=\mathcal{L}_{\rm SM}+\Delta\mathcal{L},
\label{totL}
\end{equation}
with
\begin{align}
\Delta\mathcal{L}\supset &-\frac{1}{4}C_{\mu\nu}C^{\mu\nu}-\frac{\delta}{2}C_{\mu\nu}B^{\mu\nu}+g_X J^{\mu}_X C_{\mu}-\frac{1}{2}(\partial_{\mu}\sigma+M_1 C_{\mu}+M_2 B_{\mu})^2,
\label{deltaL}
\end{align}
where $g_X$ is the gauge coupling in the hidden sector, $J_X$ is the hidden sector current and $\sigma$ is a pseudoscalar field which is absorbed in a gauge-invariant way via the Stueckelberg mechanism to give mass to the extra neutral gauge boson which we call $\gamma'$ (the dark photon). Further one may introduce matter in the hidden sector which is  
 neutral under $U(1)_Y$ but charged under $U(1)_X$~\cite{Kors:2004dx,Cheung:2007ut}.
 More generally one may have both kinetic mixing and mass mixing~\cite{Feldman:2007wj}.

The kinetic energy terms in Eqs.~(\ref{totL}) and~(\ref{deltaL}) can be diagonalized by a $GL(2,\mathbb{R})$ transformation
\beqn
\left(\begin{matrix} C^{\mu} \cr 
B^{\mu} 
\end{matrix}\right) = \left(\begin{matrix} c_{\delta} & 0 \cr 
-s_{\delta} & 1 
\end{matrix}\right)\left(\begin{matrix} C'^{\mu} \cr 
B'^{\mu} 
\end{matrix}\right), 
\label{rotation}
\eeqn  
where $c_{\delta}=1/(1-\delta^2)^{1/2}$ and $s_{\delta}=\delta/(1-\delta^2)^{1/2}$. 

In the Standard Model the neutral gauge boson sector arises from the hypercharge gauge boson
$B^\mu$ and the third component of the $SU(2)_L$ gauge field $A^\mu_a$ ($a=$ 1$-$3),
which leads to a $2\times 2$ mass squared matrix after spontaneous symmetry breaking. It contains
one massless mode, the photon, and a massive mode, the $Z$-boson. Inclusion of the Stueckelberg
gauge field $C_\mu$ enlarges the $2\times 2$ mass squared matrix of the 
neutral gauge boson sector in the standard model to  a $3\times 3$ mass squared matrix in the
Stueckelberg extended model.
Thus  after spontaneous electroweak symmetry breaking and  the Stueckelberg mass growth,
and on including the $GL(2,\mathbb{R})$ transformation 
to obtain a diagonal and a normalized kinetic energy for the gauge bosons, 
the $3\times 3$ mass squared matrix of neutral vector bosons 
 in the basis $(C'_{\mu}, B'_{\mu}, A^3_{\mu})$ is given by
\beqn
\mathcal{M}^2_V=\left(\begin{matrix}  M_1^2\kappa^2+\frac{1}{4}g^2_Y v_H^2 s^2_{\delta} & \kappa\epsilon M_1^2-\frac{1}{4}g^2_Y v_H^2 s_{\delta} & \frac{1}{4}g_Y g_2 v_H^2 s_{\delta} \cr
\kappa\epsilon M_1^2-\frac{1}{4}g^2_Y v_H^2 s_{\delta} & \epsilon^2 M_1^2+\frac{1}{4}g^2_Y v_H^2 & -\frac{1}{4}g_Y g_2 v_H^2 \cr
\frac{1}{4}g_Y g_2 v_H^2 s_{\delta} & -\frac{1}{4}g_Y g_2 v_H^2 & \frac{1}{4}g^2_2 v_H^2 \cr
\end{matrix}\right),
\label{zmassmatrix}
\eeqn
where $g_2$ is the $SU(2)_L$ gauge coupling, $\kappa=(c_{\delta}-\epsilon s_{\delta})$, $\epsilon=M_2/M_1$ and $v_H$ is the Higgs VEV. The mass-squared matrix of Eq.~(\ref{zmassmatrix}) has one zero eigenvalue which is the photon, while the other two eigenvalues are
\begin{equation}
M^2_{\pm} = \frac{1}{2}\left\{M^2_0\pm\sqrt{M_0^4-M_1^2v_H^2\Big[(\kappa^2+\epsilon^2)g_2^2+g^2_Y c^2_{\delta}\Big]}~\right\},
\label{bosons}
\end{equation}
where $M^2_0=(\kappa^2+\epsilon^2)M_1^2+\dfrac{1}{4}v_H^2(g_Y^2 c^2_{\delta}+g_2^2)$.
Here $M_-$ is identified as the $\gamma'$ boson mass, while $M_+$ as  the $Z$ boson. The diagonalization of the mass-squared matrix of Eq.~(\ref{zmassmatrix}) can be done via two orthogonal transformations, where the first is given by~\cite{Feldman:2007wj}
\beqn
\mathcal{O}=\left(\begin{matrix} 1/c_{\delta} & -s_{\delta}/c_{\delta} & 0 \cr
s_{\delta}/c_{\delta} & 1/c_{\delta} & 0 \cr
0 & 0 & 1 \cr
\end{matrix}\right),
\label{omatrix}
\eeqn
which transforms the mass matrix to $\mathcal{M'}^2_V=\mathcal{O}^{T}\mathcal{M}^2_V\mathcal{O}$, 
\beqn
\mathcal{M'}^2_V=\left(\begin{matrix}  M_1^2 & M_1^2\bar\epsilon & 0 \cr
M_1^2\bar\epsilon & M_1^2\bar\epsilon^2+\frac{1}{4}g^2_Y v_H^2 c^2_{\delta} & -\frac{1}{4}g_Y g_2 v_H^2 c_{\delta} \cr
0 & -\frac{1}{4}g_Y g_2 v_H^2 c_{\delta} & \frac{1}{4}g^2_2 v_H^2 \cr
\end{matrix}\right),
\label{zpmassmatrix}
\eeqn  
where $\bar\epsilon=\epsilon c_{\delta}-s_{\delta}$.
The gauge eigenstates of $\mathcal{M'}^2_V$ can be rotated into the corresponding mass eigenstates $(\gamma',Z,\gamma)$ using the second transformation 
such that $\mathcal{R}^T\mathcal{M'}^2_V\mathcal{R}=\text{diag}(m^2_{\gamma'},m^2_{Z},0)$
where  the rotation matrix is given by 
\beqn
\mathcal{R}=\left(\begin{matrix} \cos\psi \cos\phi-\sin\theta\sin\phi\sin\psi & \sin\psi \cos\phi+\sin\theta\sin\phi\cos\psi & -\cos\theta\sin\phi \cr
\cos\psi \sin\phi+\sin\theta\cos\phi\sin\psi & \sin\psi \sin\phi-\sin\theta\cos\phi\cos\psi & \cos\theta\cos\phi \cr
-\cos\theta \sin\psi & \cos\theta \cos\psi & \sin\theta \cr
\end{matrix}\right).
\label{rotmatrix}
\eeqn
Here the mixing angles are given by 
\begin{equation}
\tan\phi=\bar\epsilon, ~~~ \tan\theta=\frac{g_Y}{g_2}c_{\delta}\cos\phi,
\end{equation}
and
\begin{equation}
\tan2\psi\simeq\frac{2\bar\epsilon m^2_Z\sin\theta}{m^2_{\gamma'}-m^2_Z+(m^2_{\gamma'}+m^2_Z-m^2_W)\bar\epsilon^2},
\end{equation}
where $m_W=g_2 v_H/2$, $m_{\gamma'}\equiv M_-$ and $m_{Z}\equiv M_+$. Since the dark photon mixes with the SM gauge bosons, it will couple with the SM fermions and so 
\begin{equation}
\mathcal{L}_{\rm SM}=\frac{g_2}{2\cos\theta}\bar\psi_f\gamma^{\mu}\Big[(v_f-\gamma_5 a_f)Z_{\mu}+(v'_f-\gamma_5 a'_f)A^{\gamma'}_{\mu}\Big]\psi_f+e\bar\psi_f\gamma^{\mu}Q_f A_{\mu}\psi_f,
\label{SMLag}
\end{equation}
where $f$ runs over all SM fermions and the vector and axial couplings are given by
\begin{equation}
\begin{aligned}
v_f&=\cos\psi[(1-\bar\epsilon\tan\psi\sin\theta)T_{3f}-2\sin^2\theta(1-\bar\epsilon\csc\theta\tan\psi)Q_f],\\
a_f&=\cos\psi(1-\bar\epsilon\tan\psi\sin\theta)T_{3f}, \\
v'_f&=-\cos\psi[(\tan\psi+\bar\epsilon\sin\theta)T_{3f}-2\sin^2\theta(\bar\epsilon\csc\theta+\tan\psi)Q_f],\\
a'_f&=-\cos\psi(\tan\psi+\bar\epsilon\sin\theta)T_{3f}.
\end{aligned}
\end{equation}
Here $T_{3f}$ is the third component of the isospin and $Q_f$ is the electric charge. 

We assume that the hidden sector where $C^\mu$ resides contains two mass degenerate 
 Dirac fermions $D_1$ and $D_2$ with the common mass $\mu$ 
 which, however, carry different charges $Q_1$ and $Q_2$ under 
 the $U(1)_X$ gauge group.  The interaction Lagrangian
for the hidden sector is then given by
\begin{align}
 \mathcal{L}^{\rm int}_D=& -g^{\gamma'}_X Q_1 \bar D_1\gamma^\mu D_1 A_\mu^{\gamma'} - g^{\gamma'}_X Q_2 \bar D_2\gamma^\mu D_2  A_\mu^{\gamma'}-g^{Z}_X Q_1 \bar D_1\gamma^\mu D_1 Z_\mu-g^{Z}_X Q_2 \bar D_2\gamma^\mu D_2  Z_\mu,
\label{d1d2-1}
\end{align}
with $g^{\gamma'}_X =g_X(\mathcal{R}_{11}-s_{\delta}\mathcal{R}_{21})$ and $g^{Z}_X =g_X(\mathcal{R}_{12}-s_{\delta}\mathcal{R}_{22})$, where $\mathcal{R}_{ij}$ are elements of the matrix in Eq.~(\ref{rotmatrix}). 
To generate inelastic scattering we need to split the masses of the $D$-fermions. To this end we
add a $U(1)_X$ gauge violating mass terms $\Delta\mu (\bar D_1D_2 + \text{h.c.})$ so that the Lagrangian for the
 $(D_1, D_2)$ mass terms  is  given by 
 \begin{align}
  \mathcal{L}^{\rm mass}_D = 
 - \mu (\bar D_1 D_1 + \bar D_2 D_2)-
 \Delta \mu (\bar D_1 D_2 + \bar D_2 D_1),
 \label{d1d2-2}
 \end{align}
We can now go to the mass diagonal basis with Dirac fermions $D_1'$ with mass $m_1=\mu-\Delta \mu$
and $D_2'$ with mass $m_2= \mu+\Delta \mu$ and we assume $m_2>m_1$ so $D_2'$ is the heavier
of the two dark fermions. In this basis, Eq.~(\ref{d1d2-1}) takes the form
\begin{equation}
\begin{aligned}
-\mathcal{L}^{\rm int}_D&= \frac{1}{2}g^{\gamma'}_X (Q_1+ Q_2) \left( \bar D_1'\gamma^\mu D_1' 
+ \bar D_2'\gamma^\mu D_2'\right)A_\mu^{\gamma'} \\
&+\frac{1}{2}  g^{\gamma'}_X (Q_1-Q_2) (\bar D_1' \gamma^{\mu}D_2' + \bar D_2' \gamma^{\mu}D_1')A_\mu^{\gamma'} \\
& +\frac{1}{2} g^{Z}_X (Q_1 + Q_2) \left( \bar D_1'\gamma^\mu D_1'
+ \bar D_2'\gamma^\mu D_2'\right) Z_\mu \\
&+ \frac{1}{2} g^{Z}_X (Q_1-Q_2) (\bar D_1' \gamma^{\mu}D_2' + \bar D_2' \gamma^{\mu}D_1')Z_\mu.
\end{aligned}
\label{d1d2-2} 
\end{equation}
  From Eq.~(\ref{SMLag}) and Eq.~(\ref{d1d2-2}) we note that the dark photon has couplings
  with both the visible sector and the hidden sector. Thus from Eq.~(\ref{SMLag}) we find that 
  the dark photon couples with quarks and leptons in the visible sector while from    
  Eq.~(\ref{d1d2-2}) we find that the dark photon has 
   couplings with  $D_1', D_2'$ in the dark sector. 
These couplings allow for  an inelastic scattering 
   process to occur where a dark fermion hits a bound electron in a xenon atom producing, in an exothermic process, a recoil electron with excess energy, i.e.,  
\begin{align}
e+ D_2'\to e'+ D_1', 
\end{align}
where the final electron receives an extra boost in energy from the mass difference $\Delta m= m_2-m_1
= 2 \Delta \mu$.

\section{Dark matter relic density\label{sec:dm}}
Since the hidden sector matter has feeble couplings with the Standard Model particles, they are
never in thermal equilibrium with the visible sector and the usual freeze-out analysis for the
computation of the relic density does not apply. However, the hidden sector particles can be 
produced via the annihilation  of the Standard Model particles  into dark photons and 
dark fermions via these feeble interactions and the computation of the relic density in this 
case is computed using the freeze-in mechanism~\cite{Hall:2009bx}.
 Within the dark sector itself the 
 dark fermions and the dark photons interact via normal size interactions
 and are in thermal equilibrium up to a certain freeze-out temperature $T_f$ via 
 the process $D\bar{D}\to\gamma'\gamma'$. However, below the dark sector freeze-out temperature
 $T_f$, the dark fermions and the dark photons decouple and the 
 dark photons decay to the visible sector much before the big bang nucleosynthesis, which leaves the dark fermions as the only DM candidates. In our model we assume that the visible and hidden sectors have the same temperature. One can consider sectors with different temperatures in the early universe, but it has been shown that the two sectors will eventually thermalize and that the effect on the relic density is minimal~\cite{Aboubrahim:2020lnr}. 

For calculating the DM relic density of $D'_1$ and $D'_2$, we assume $m_{1}\simeq m_{2}\simeq m_D$ and write only one Boltzmann equation for the dark fermion. In this limit, we have
\begin{equation}
\frac{dY_D}{dx}\approx -1.32 M_{\rm Pl}\frac{h_{\rm eff}(T)}{g^{1/2}_{\rm eff}(T)}\frac{m_D}{x^2}\left(-\langle\sigma v\rangle_{D\bar{D}\to i\bar{i}} Y_D^{\rm eq^2} +\langle\sigma v\rangle_{D\bar{D}\to\gamma'\gamma'} Y^2_D\right),
\label{yield}
\end{equation} 
where $Y_D=n/s$ is the comoving number density (or yield) of DM, $h_{\rm eff}$ and $g_{\rm eff}$ are the entropy and energy density numbers of degrees of freedom,  $M_{\rm Pl}$ is the reduced Planck mass
($M_{\rm Pl} \sim 2.4 \times 10^{16}$ GeV) and $x=m_D/T$. The first term on the right-hand-side of Eq.~(\ref{yield}) is the production of dark matter particles via the freeze-in mechanism and the second term produces  DM depletion. Here the thermally averaged cross-section is given by
\begin{equation}
\langle\sigma v\rangle^{D\bar{D}\to ab}(x)=\frac{x}{8 m^5_D K^2_2(x)}\int_{4m_D^2}^{\infty} ds ~\sigma(s) \sqrt{s}\, (s-4m_D^2)K_1\left(\frac{\sqrt{s}}{m_D}x\right),
\end{equation}
while the equilibrium yield is given by
\begin{equation}
Y_D^{\rm eq}(x)=\frac{45}{4\pi^4}\frac{g_D}{h_{\rm eff}(T)} x^2 K_2(x).
\end{equation}
Here $g_D$ is the dark fermion number of degrees of freedom, $K_1$ and $K_2$ are the modified second order Bessel functions of degree one and two. The Boltzmann equation Eq.~(\ref{yield}) 
is solved numerically to determine the yield at the present time $Y_{\infty}$ which gives us the relic density
\begin{equation}
\Omega h^2=\frac{m_D Y_{\infty}s_0 h^2}{\rho_c},
\end{equation}
where $s_0$ is today's entropy density, $\rho_c$ is the critical density and $h=0.678$ denotes the present Hubble expansion rate in units of 100 km s$^{-1}$ Mpc$^{-1}$. 

We select three benchmarks with different masses for the dark photon and for the dark 
 fermions. The benchmarks are listed in Table~\ref{tab1} along with the couplings, the relic density and the inelastic DM-electron scattering cross-section. In the analysis, we set $\epsilon=0$, i.e. we assume no mass mixing.

\begin{table}[H]
\begin{center}
\begin{tabulary}{1.00\textwidth}{l|CCCCCCCCC}
\hline\hline\rule{0pt}{3ex}
Model & $m_D$ (GeV) & $m_{\gamma'}$ (MeV) & $g_X$ & $\delta$ & $\Omega h^2$ & $\bar\sigma_e$ (cm$^2$)  \\
\hline\rule{0pt}{3ex}
\!\!(a) & 1.00 & 55 & 0.040 & $4.0\times 10^{-5}$ & 0.125 & $2.80\times 10^{-44}$ \\
(b) & 0.50 & 60  & 0.025 & $6.0\times 10^{-5}$ & 0.121 & $1.75\times 10^{-44}$ \\
(c) & 0.30 & 300  & 0.055 & $7.5\times 10^{-4}$ & 0.116 & $1.19\times 10^{-44}$ \\
\hline
\end{tabulary}\end{center}
\caption{Input parameters, the relic density $\Omega h^2$ and the inelastic DM-electron scattering cross  section $\bar\sigma_e$ for the benchmarks used in this analysis.}
\label{tab1}
\end{table}

\begin{figure}[H]
 \centering
   \includegraphics[width=0.49\textwidth]{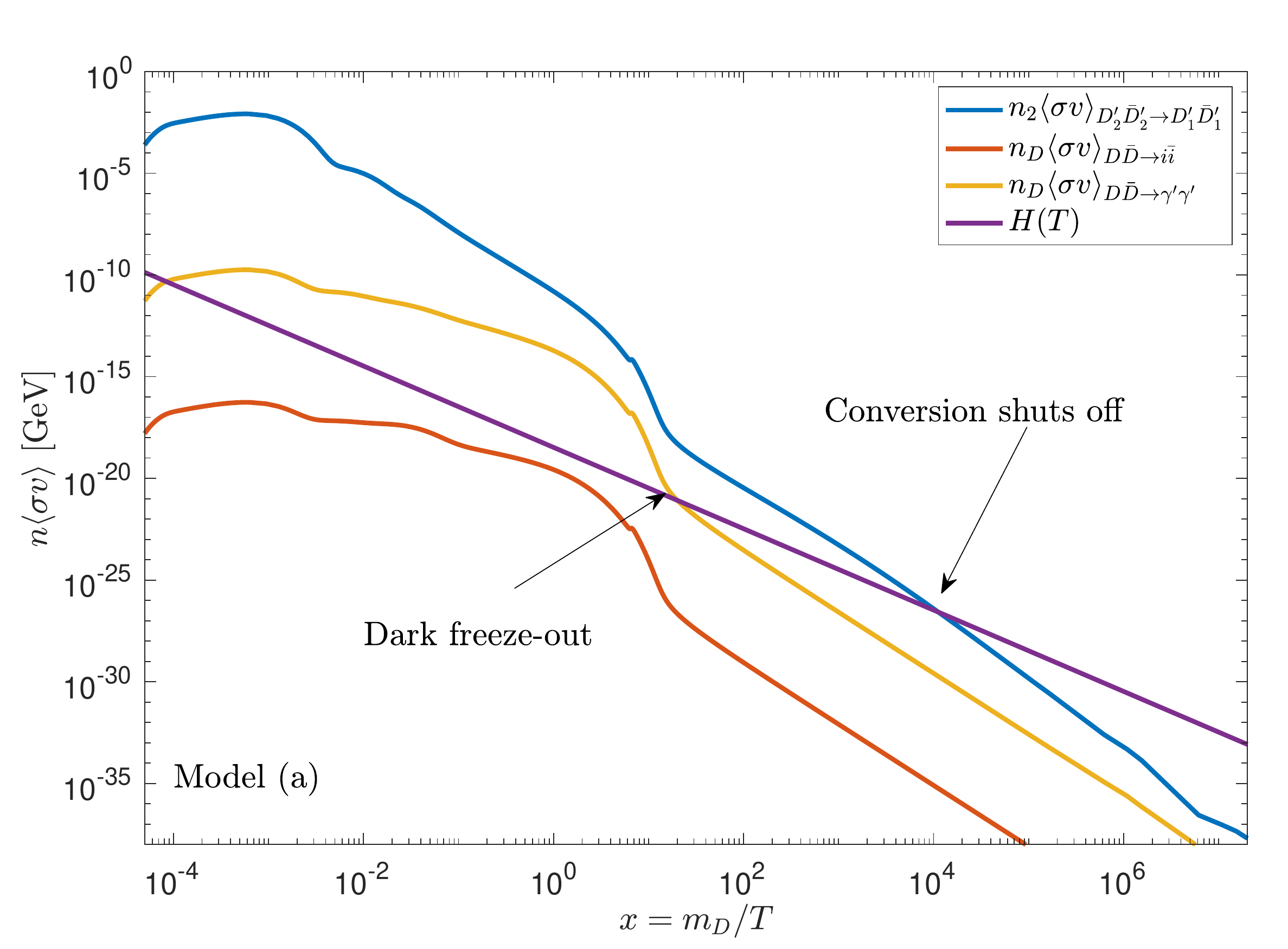}
   \includegraphics[width=0.49\textwidth]{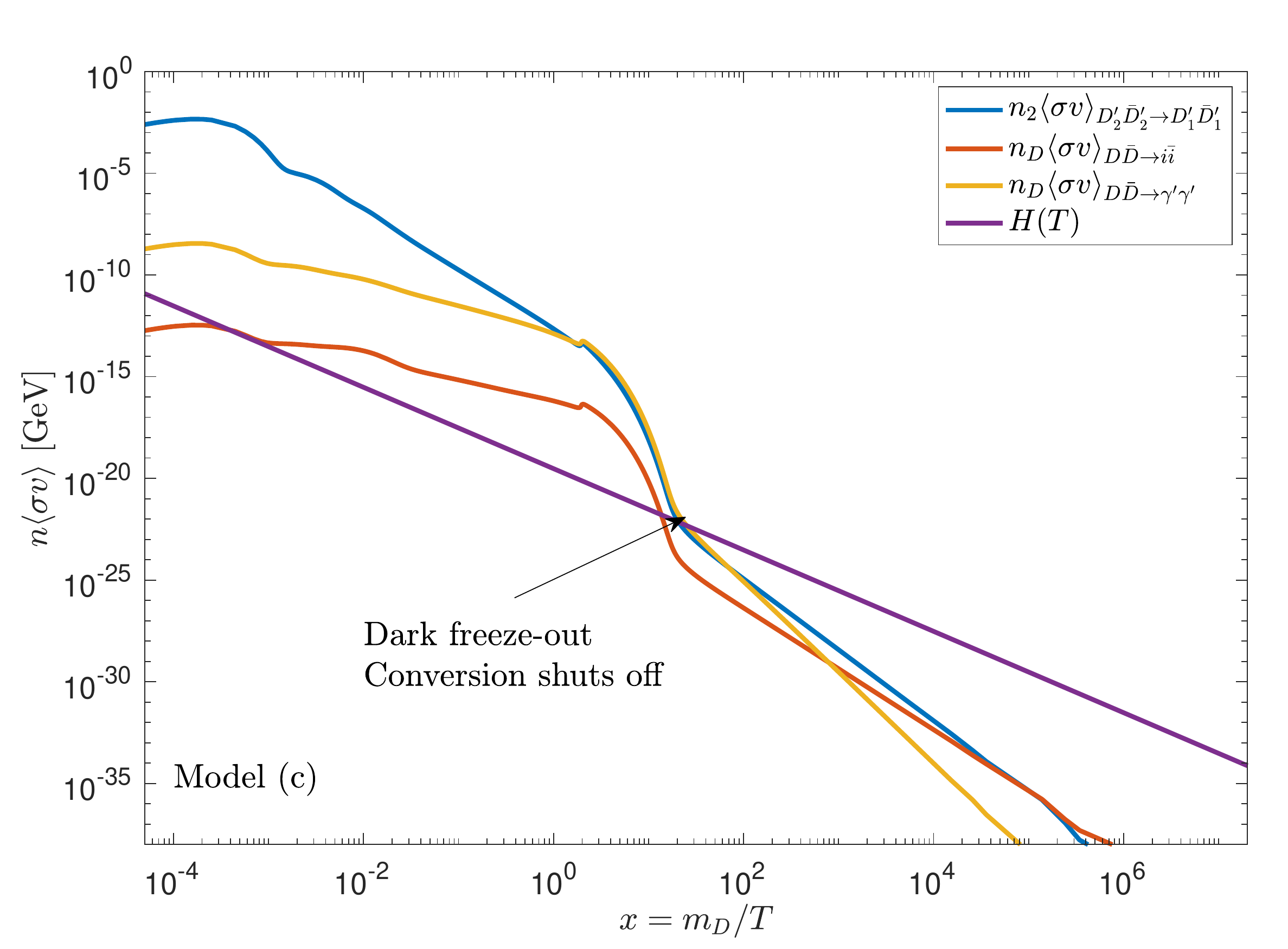}
   \caption{A plot of $n\langle\sigma v\rangle$ and the Hubble parameter $H(T)$ as a function of 
   $x=m_D/T$ for benchmark (a) (left panel) and (c) (right panel). Dark freeze-out sets in before the DM conversion process shuts off for model (a). The analysis of  model (b) is similar to that of model
   (a),  while both processes occur nearly at the same temperature for model (c).}
	\label{fig1}
\end{figure}

The DM relic density satisfies, within theoretical uncertainties, the experimental value from the Planck Collaboration~\cite{Aghanim:2018eyx}
\begin{equation}
(\Omega h^2)_{\rm Planck}=0.1198\pm 0.0012.
\end{equation}
Following the dark freeze-out of DM in the hidden sector, conversion processes $D'_2\bar{D}'_2\longleftrightarrow D'_1\bar{D}'_1$ remain active. The ratio of the number densities of $D'_1$ and $D'_2$ is Boltzmann suppressed, i.e.,  $n_{2}/n_{1}\sim\exp(-\Delta m/T_c)$,
where  $T_c$ is the temperature below which conversion processes shut off. One can safely assume that $n_{2}\sim n_{1}$ as long as $T_c\gg \Delta m$. Focusing on the process $D'_2\bar{D}'_2\longrightarrow D'_1\bar{D}'_1$, we determine $T_c$ for which
\begin{equation}
n_{2}\langle\sigma v\rangle_{D'_2\bar{D}'_2\longrightarrow D'_1\bar{D}'_1}/H\sim 1,
\end{equation}
where $H$ is the Hubble parameter.
In Fig.~\ref{fig1} we plot $n\langle\sigma v\rangle$ versus $x$ for benchmarks (a) (left panel) and (c) (right panel). The figure shows the conversion process (blue) along with two other processes, $D\bar{D}\to i\bar{i}$ (red) and $D\bar{D}\to\gamma'\gamma'$ (yellow) and the Hubble parameter (purple). For benchmark (a), one finds that for the value of the kinetic mixing chosen the 
DM remains out of equilibrium with the SM for the entire $x$ range, while increasing the kinetic mixing causes such a process to enter in equilibrium for a range of $x$ as shown for benchmark (c). This process, however, decouples before other decoupling processes. 

As also shown in Fig.~\ref{fig1}, dark freeze-out occurs for $x\sim 20$ for all benchmarks. Turning to the conversion process, for benchmark (a), such a process shuts off at $x\sim 1.12\times 10^4$, which for the DM mass of benchmark (a) corresponds to $T_c\sim 90$ keV. The mass gap of interest here is $\Delta m\sim 2.8$ keV which makes the number densities of $D'_1$ and $D'_2$ almost equal. One can see that the conversion process shuts off much later than the other ones. This is also observed for benchmark (b). Unlike (a) and (b), benchmark (c) shows the freeze-out and conversion process shutting off occur at the same $x\sim 20$ which corresponds to a temperature much larger than $\Delta m$. Thus for practical purposes hereafter, we assume that the DM density is divided equally between $D'_1$ and $D'_2$.

After the conversion process terminates, the dark fermion $D_2'$ can decay to  $D_1'$. The only decay channel would be $D_2'\to D_1'\bar \nu\nu$, which is further suppressed by the dark photon coupling to the neutrinos which is proportional to the kinetic mixing, and further the decay is phase-space suppressed because
of the small mass gap $\Delta m\sim\mathcal{O}$(keV).  The total 3-body decay width is (including three neutrino generations)
\begin{equation}
\Gamma_{D_2'\to D_1'\nu\bar\nu}=\frac{x_{\nu}^2}{256\pi^3 m^4_{\gamma'}}\left[\frac{f(m_1,m_2)}{m_2^3}+24 m_1^3(m_1^2+m_1m_2+m^2_2)\log\left(\frac{m_2}{m_1}\right)\right],
\label{widthtot}
\end{equation}
where
\begin{equation}
f(m_1,m_2)=(m_2^2-m_1^2)(m_1^6-2m_2 m_1^5-7m_2^2 m_1^4-20m_2^3 m_1^3-7m_2^4 m_1^2-2m_2^5 m_1+m_2^6).
\end{equation}
Since $m_1=m_2-\Delta m$ and expanding in $\Delta m$, we get to lowest order in $\Delta m$
\begin{equation}
\Gamma_{D_2'\to D_1'\nu\bar\nu}\simeq\frac{x_{\nu}^2(\Delta m)^5}{40\pi^3 m^4_{\gamma'}},
\end{equation}
where for a small gauge kinetic mixing, $g_X^{\gamma'}\approx g_X$ and
\begin{equation}
x_{\nu}\sim g_X g_Y(Q_1-Q_2)\left(\frac{m_{\gamma'}}{m_Z}\right)\delta.
\end{equation}
In this work we are interested in $\Delta m\sim 3$ keV and $\delta\sim 10^{-5}$ which results in a decay lifetime of $D_2'$ order $10^{13}$ years which means that $D_2'$  is stable over the lifetime of the universe. Thus, in this model dark matter is constituted of two dark fermions with essentially
degenerate masses which are of order 1 GeV.

\section{DM-electron scattering cross-section \label{sec:e-dm} }

A DM particle can undergo an elastic scattering with a bound electron in a xenon atom, but such a scattering can deliver only a few eV to the electron which is not sufficient to explain the Xenon-1T excess.
However, an inelastic exothermic down-scattering can impart a recoil energy to the electron equivalent to the mass difference between the incoming and outgoing DM particles. 
The model considered here  allows for the desired small mass splitting between the two Dirac fermions $D'_1$ and $D'_2$ so that the heavier fermion $D'_2$ down-scatters to $D'_1$.
Next, we compute the inelastic scattering cross-section of the process described $D_2'(\vec{p}_1)+e(\vec{p}_2)\to D_1'(\vec{p}_3)+e'(\vec{p}_4)$. Assuming the dark photon mass is much greater than the momentum transfer, the averaged matrix element squared for this process
 is given by
\begin{align}
\overline{|\mathcal{M}|^2}&=\frac{2\bar{g}_X^2 g_2^2}{m^4_{\gamma'}\cos^2\theta}\Bigg\{\frac{1}{2}(a_f^{\prime2}-v_f^{\prime2})\Big[(m_1-m_2)^2-(t+2m_1m_2)\Big]m^2_e +\frac{1}{4}(v_f^{\prime2}+a_f^{\prime2})\Big[(m_2^2+m_e^2-u) \nonumber \\
&\times (m_1^2+m_e^2-u)+(s-m_1^2-m_e^2)(s-m_2^2-m_e^2)-2m_1m_2(2m_e^2-t)\Big]\Bigg\},
\end{align} 
where $\bar{g}_X=\frac{1}{2} g_X(Q_1-Q_2)$ and $s,t,u$ are the Mandelstam variables.
The directional matrix element for free electron-DM scattering is given by
\begin{equation}
\overline{|\mathcal{M}(\vec{q})|^2}=\overline{|\mathcal{M}(q)|^2}\times |F_{DM}(q)|^2,
\end{equation}
where the form factor $F_{DM}(q)$ can be taken as 1 for a small momentum transfer. The electron-DM scattering differential cross-section is given by
\begin{align}
\frac{d\bar{\sigma}_e}{d\Omega}&=\frac{1}{64\pi^2 s}\frac{|\vec{p}_3|}{|\vec{p}_1|}\overline{|\mathcal{M}(\vec{q})|^2}.
\end{align}
Keeping the velocity dependence in the Mandelstam variables, we have
\begin{equation}
t=-q^2\simeq -\Delta m\left(1-\sqrt{\frac{2m_e v^2}{\Delta m}}\cos\theta_{\rm CM}\right),~~~s\sim (m_2+m_e)^2\left(1+\frac{\mu_{De}}{m_2+m_e}v^2\right),
\end{equation}
and integrating over $\theta_{\rm CM}$, the scattering angle in the CM frame, and $\phi$, we get for the DM-e scattering cross-section
\begin{equation}
\bar{\sigma}_e\simeq\frac{\bar{g}_X^2 g_2^2}{16\pi m^4_{\gamma'}\cos^2\theta}\left(\frac{4\mu^2_{De}}{1+\frac{\mu_{De}}{m_2+m_e}v^2}\right)\left[v^{\prime 2}_f+(a_f^{\prime 2}+v^{\prime 2}_f) v^2\right],
\end{equation} 
where $\mu_{De}=\frac{m_2 m_e}{m_2+m_e}$ is the dark matter-electron reduced mass. For $v\sim 10^{-3}$, one can discard the velocity-dependent piece and  get
\begin{equation}
\bar{\sigma}_e\simeq\frac{\bar{g}_X^2 g_2^2}{4\pi\cos^2\theta}\frac{\mu^2_{De}}{m^4_{\gamma'}}v_f^{\prime 2}.
\end{equation}
The values of the cross-section for the three benchmarks are shown in Table~\ref{tab1}. One finds that the cross-section depends on the gauge coupling in the dark sector and on kinetic mixing which enters in the expression of $v'_f$. Such quantities are constrained by experiments which we discuss in section~\ref{sec:fit}.

\section{Detection rate at Xenon-1T\label{sec:detection}}
\begin{figure}[H]
 \centering
   \includegraphics[width=0.75\textwidth]{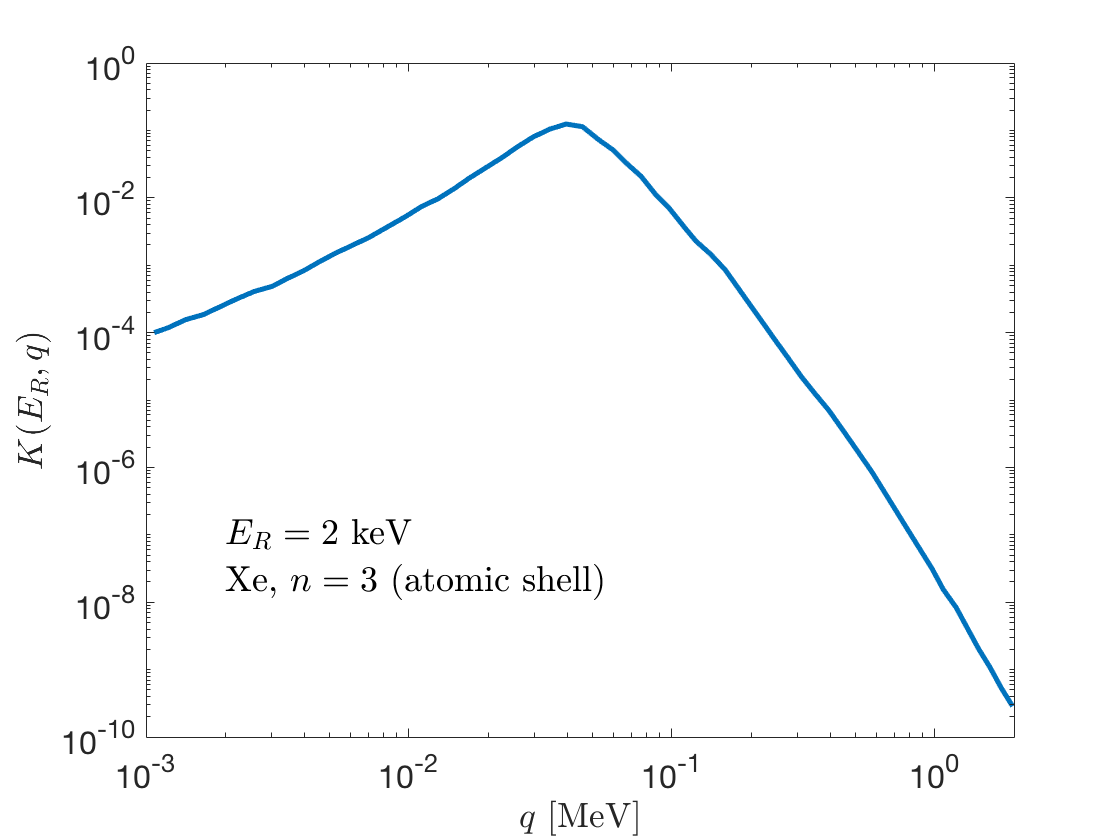}
   \caption{The atomic ionization factor $K$ summed over all possible atomic electrons dominated by $n=3$ for Xe at electron recoil energy   
   $E_R=2$ keV. The plot is a function of the momentum transfer $q$, taken from Ref.~\cite{Roberts:2019chv}.}
	\label{fig2}
\end{figure}

We give here a quantitative analysis of the excess seen in the event rate in the Xenon-1T experiment
arising from $D_2'$ with mass $m_2$ scattering inelastically off an electron into $D_1'$ with mass $m_1$   delivering a recoil energy $E_R$ to the electron. Energy conservation for this process gives
\begin{equation}
\frac{q^2}{2m_2}-vq\cos\eta=\Delta m-E_R,
\end{equation}
where $\eta$ is the angle between the incoming $D_2'$ 
momentum and the momentum transfer $\vec{q}$. Taking $m_1\approx m_2$, the range of momentum transfer is given by
\begin{equation}
    q_{\pm}= 
\begin{cases}
    m_2 v\pm\sqrt{m_2^2 v^2-2m_2(E_R-\Delta m)},& \text{for } E_R>\Delta m\\
    \pm m_2 v+\sqrt{m_2^2 v^2-2m_2(E_R-\Delta m)}, & \text{for } E_R<\Delta m.
\end{cases}
\end{equation}
The recoil energy can be expressed in terms of the mass difference and in the limit $\Delta m\ll m_e\ll m_2$, we have~\cite{Lee:2020wmh}
\begin{equation}
E_R\simeq \Delta m\left(1-\sqrt{\frac{2m_e v^2}{\Delta m}}\cos\theta_{\rm CM}\right),
\label{er}
\end{equation}
with $q^2\simeq 2m_e E_R$, where $\theta_{\rm CM}$ is the scattering angle in the CM frame. The velocity-averaged differential cross-section for inelastic DM scattering is
\begin{equation}
\frac{d\langle\sigma v\rangle}{d E_R}=\int_{v_{\rm min}}^{v_{\rm max}}\frac{f(v)}{v}dv \frac{\bar\sigma_e}{2m_e}\int_{q_{-}}^{q_{+}} dq\,a_0^2 q K(E_R,q),
\label{K-int}
\end{equation}
where the Bohr radius $a_0=1/(\alpha_{\rm em} m_e)$ $(\alpha_{\rm em}\simeq 1/137)$ and $K(E_R,q)$ is the atomic factorization factor (shown in Fig.~\ref{fig2}) and $f(v)$ is the standard Boltzmann velocity distribution after integrating the angular part. In Eq.~(\ref{K-int}) the integral on $dq$, i.e., 
\begin{equation}
K'(E_R)\equiv \int_{q_{-}}^{q_{+}} dq\,a_0^2 q K(E_R,q),
\end{equation}
can be directly evaluated  by using Fig.~\ref{fig2}.
 
From Eq.~(\ref{er}), the range of the electron recoil energy is
\begin{equation}
E_R^{\pm}\simeq \Delta m\left(1\pm\sqrt{\frac{2m_e v^2}{\Delta m}}\right),
\end{equation} 
and the differential cross-section in this range becomes
\begin{equation}
\frac{d\langle\sigma v\rangle}{d E_R}=\frac{\bar\sigma_e}{2m_e} K'(E_R) \int_{0}^{v_{\rm max}}\frac{f(v)}{v}dv ~\Theta(E_R-E_R^{-})\Theta(E_R^{+}-E_R),
\end{equation}
where for $E_R^+-E_R^-\ll E_R^{\pm}$, and where $\Theta(E_R-E_R^{-})\Theta(E_R^{+}-E_R)\simeq (E_R^+-E_R^-)\delta(E_R-\Delta m)$. Thus we have
\begin{equation}
\frac{d\langle\sigma v\rangle}{d E_R}=\sqrt{\frac{2\Delta m}{m_e}}\bar\sigma_e K'(E_R)\delta(E_R-\Delta m) \int_{0}^{v_{\rm max}}f(v) dv.
\end{equation}
Note that for $v_{\rm max}=\sqrt{2\Delta m/m_e}\gg v_0$ (the most probable velocity), we get $\int_{0}^{v_{\rm max}}f(v) dv\simeq 1$. In practice, the electron recoil energy is not manifested as a Dirac delta function but rather smeared by the detector resolution. This can be modeled by~\cite{Aprile:2020tmw}
\begin{equation}
\sigma_r=a\sqrt{E_R}+b ~E_R,
\end{equation}
with $a=(0.310\pm 0.004)\sqrt{\text{keV}}$ and $b=0.0037\pm 0.0003$. We assume the resolution function is a Gaussian of the form
\begin{equation}
R_S(E,E_R)=\frac{1}{\sqrt{2\pi}\sigma_r}\exp\left[-\frac{(E-E_R)^2}{\sigma^2_r}\right]~\alpha(E),
\end{equation}
where $\alpha(E)$ is the efficiency given in Fig. 2 of Ref.~\cite{Aprile:2020tmw} which we take to be 0.8 for our purposes. As a result, the DM detection rate is 
\begin{align}
\frac{dR}{dE}&=n_{\rm Xe}\frac{\rho_2}{m_2}\int\frac{d\langle\sigma v\rangle}{d E_R}R_S(E,E_R)dE_R \nonumber \\
&=n_{\rm Xe}~\rho_2\sqrt{\frac{2\Delta m}{m_e}}\frac{\bar\sigma_e}{m_2} K'(\Delta m)R_S(E,\Delta m),
\end{align}
where $n_{\rm Xe}\simeq 4.2\times 10^{27}/$ton is the number of xenon atoms in the detector and $\rho_2\simeq 0.15$ GeV/cm$^3$ assuming that $D_2'$ makes half the amount of the observed relic density. 
At the recoil energy of interest, $E_R\simeq\Delta m\simeq 2.5$ keV and $K'(\Delta m)\simeq 19.4$. The event detection rate becomes
\begin{equation}
\frac{dR}{dE}\simeq (1.5\times 10^{45}~\text{GeV/cm}^2)\frac{\bar\sigma_e}{m_2}R_S(E,\Delta m),
\label{rate}
\end{equation}
which has units of $(\text{t}\cdot\text{yr}\cdot\text{keV})^{-1}$.

\section{Constraints and fit to Xenon-1T data}\label{sec:fit}

Using Eq.~(\ref{rate}) we attempt to fit the theory predictions of the model based on the benchmarks (a), (b) and (c) of Table~\ref{tab1} to the Xenon-1T data. But before doing so, let us discuss the  stringent experimental constraints that must be satisfied.

\begin{figure}[H]
\centering
\includegraphics[width=0.32\textwidth]{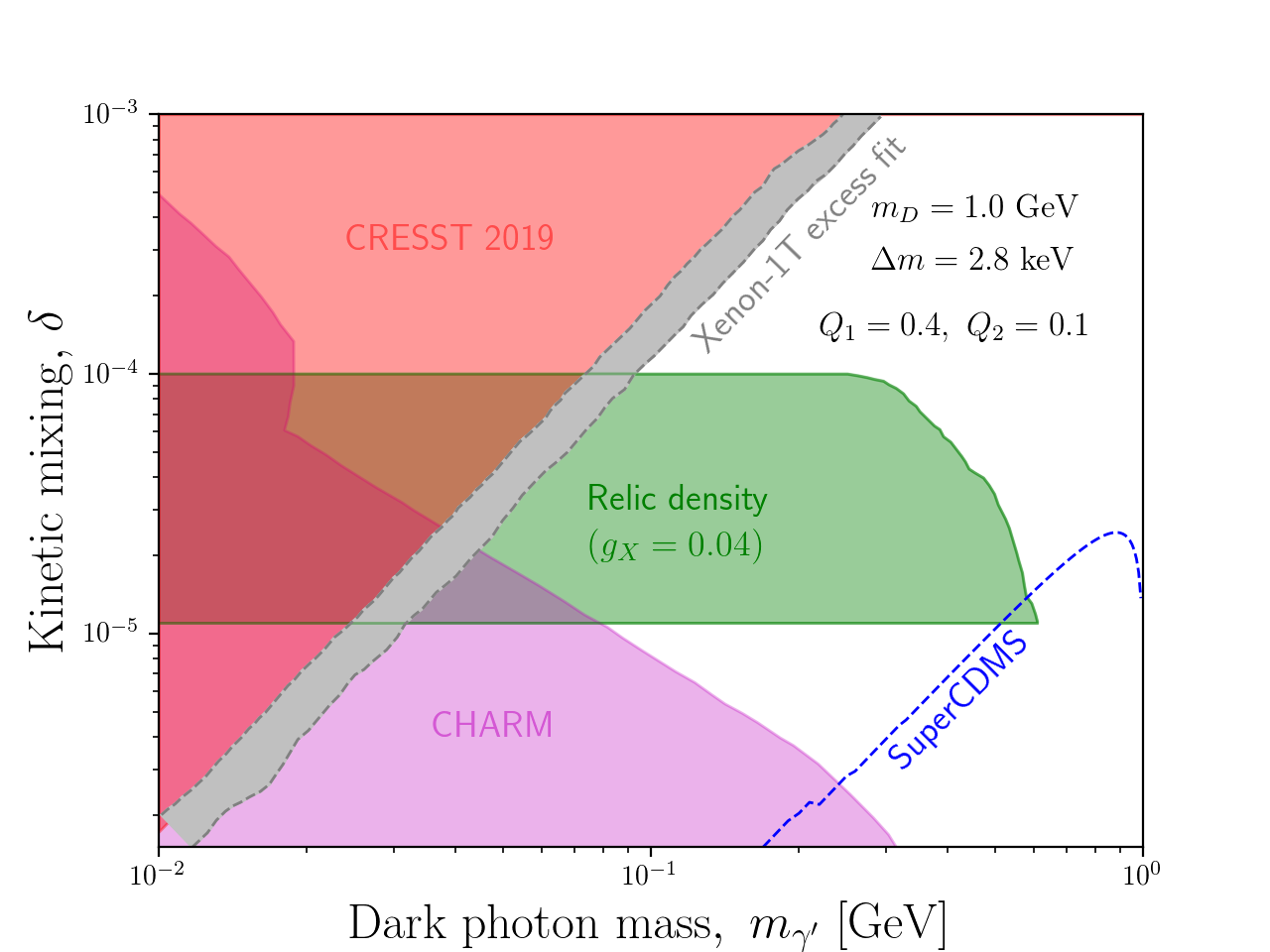}
\includegraphics[width=0.32\textwidth]{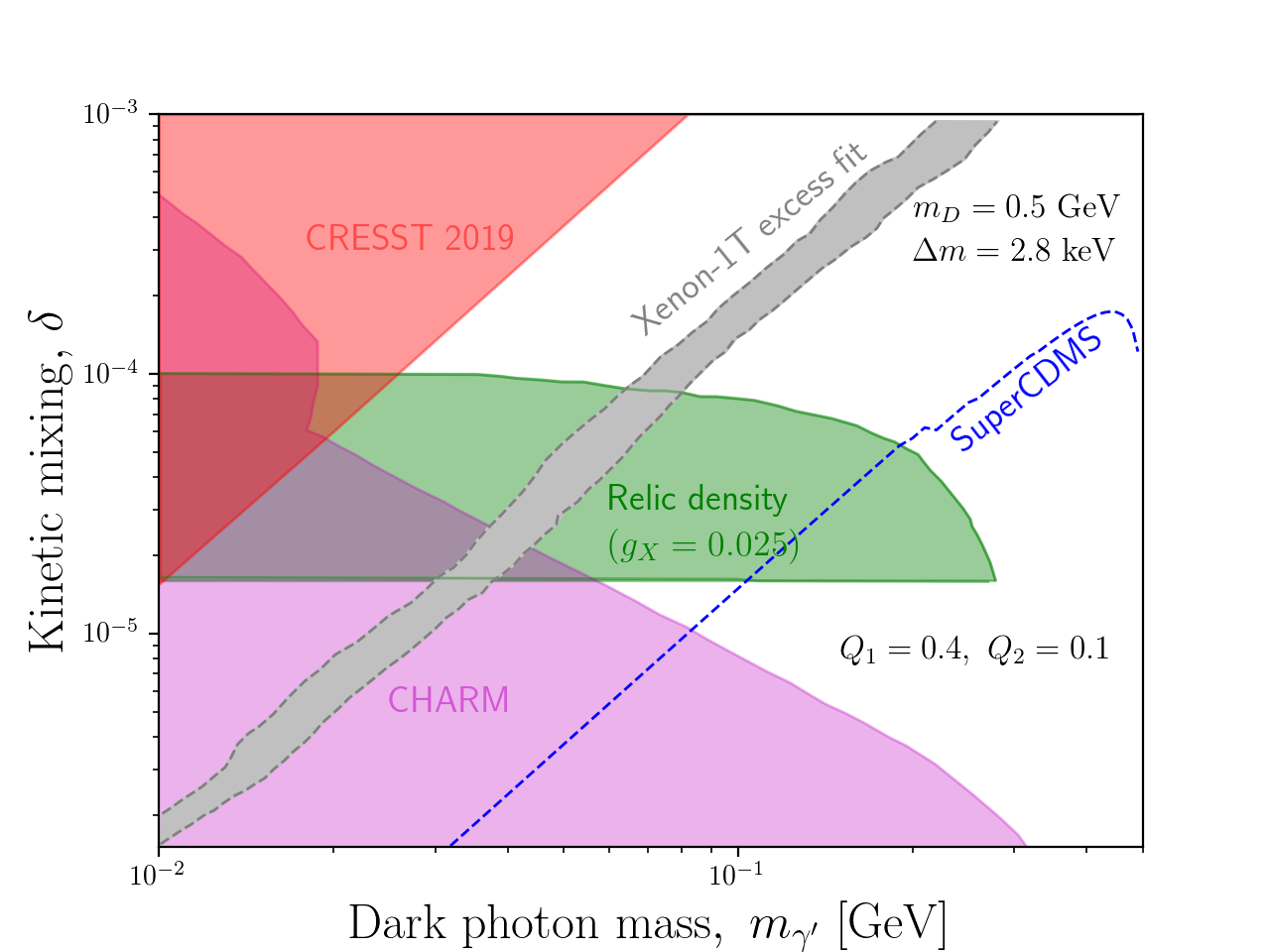}
\includegraphics[width=0.32\textwidth]{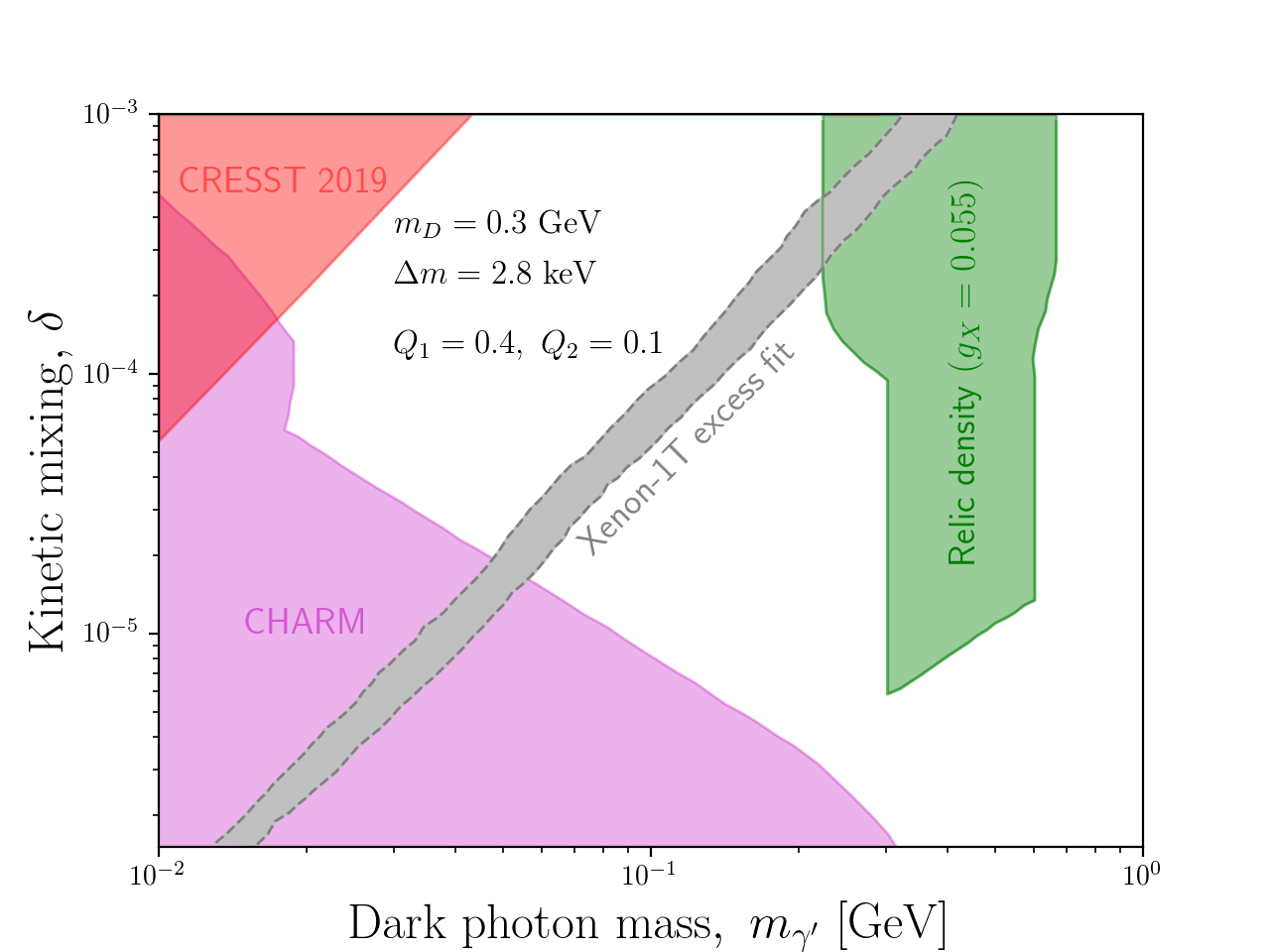}
\\
\includegraphics[width=0.32\textwidth]{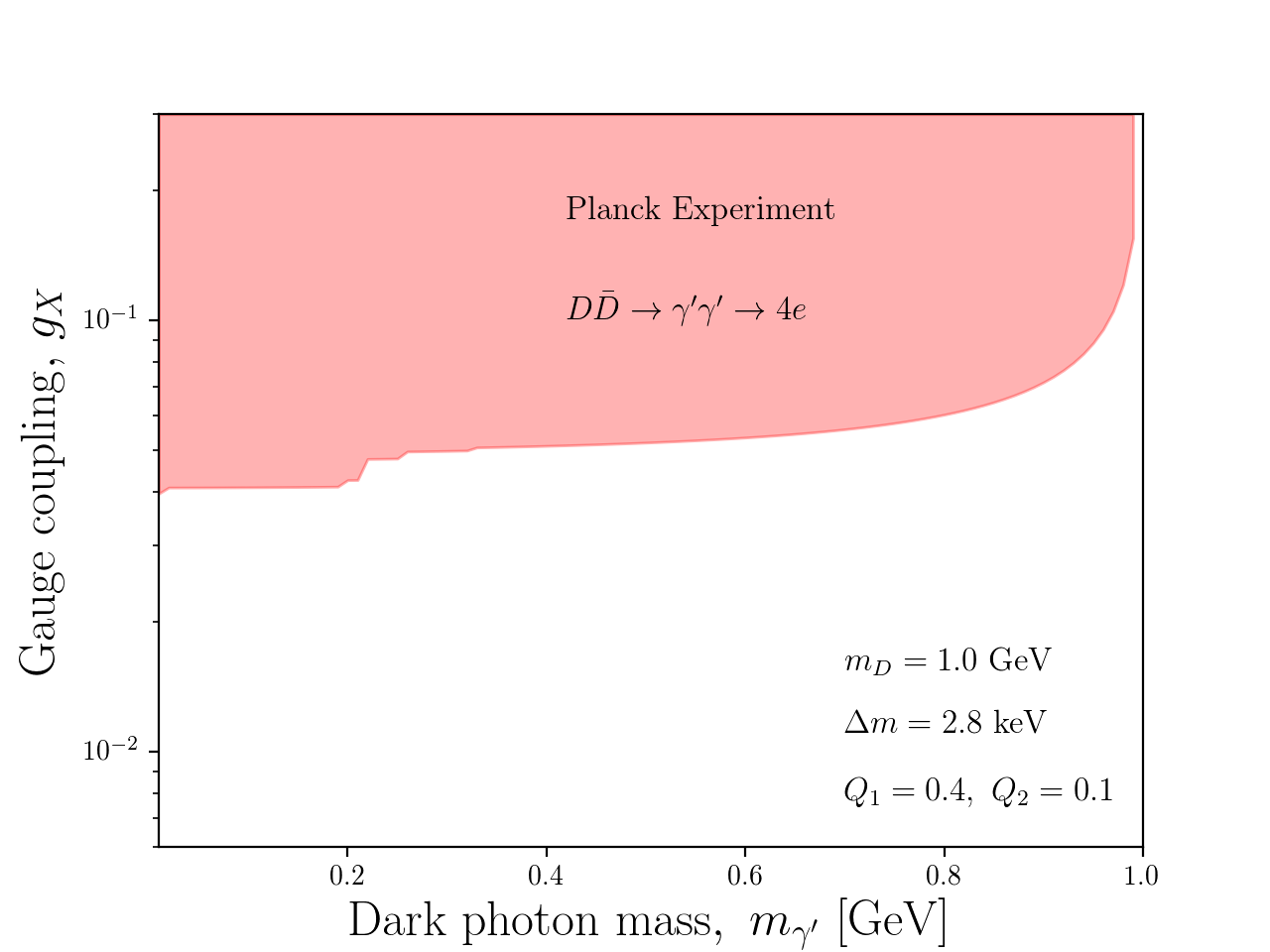}     
\includegraphics[width=0.32\textwidth]{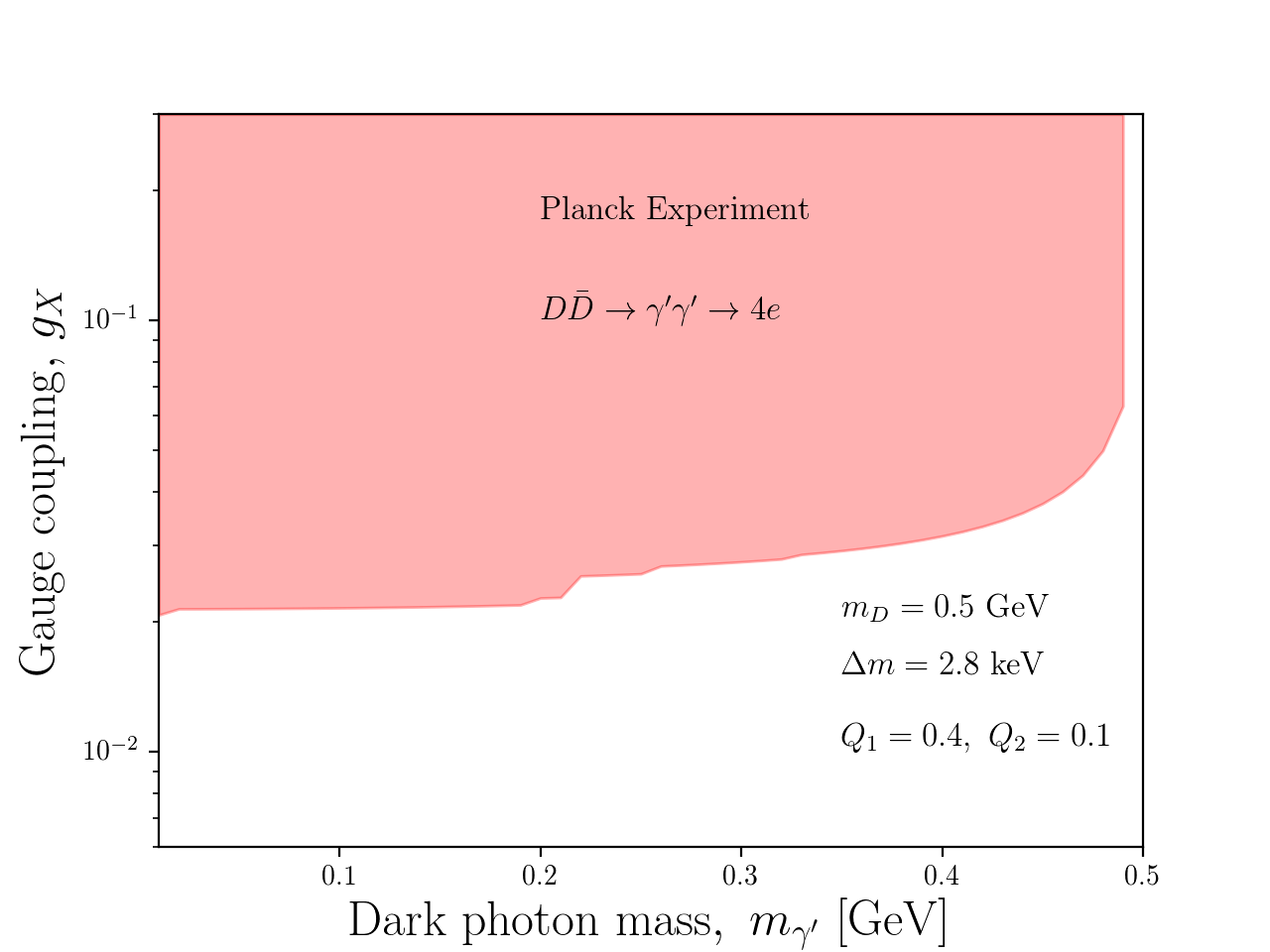}
\includegraphics[width=0.32\textwidth]{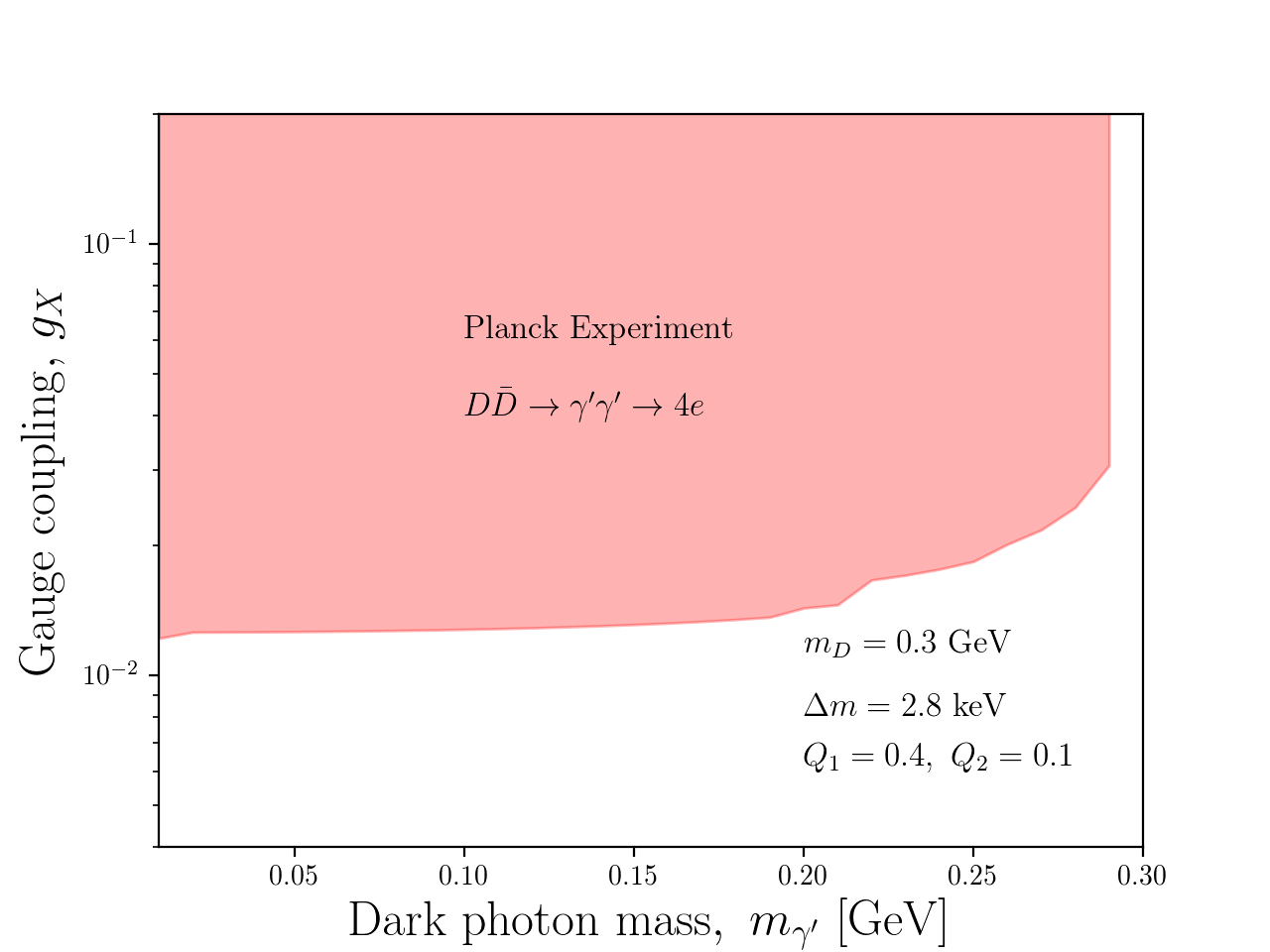}
\caption{Top panels:
Exclusion limits for benchmarks (a), (b) and (c) of Table~\ref{tab1} in the kinetic mixing-dark photon mass plane specific to our model with constraints from the CRESST 2019 DM-nucleon scattering cross-section and the projected SuperCDMS limit. Also shown is the dark photon constraint from CHARM. The green patches show the regions where the DM relic density is satisfied and the grey stripes indicate the region producing the fit to the Xenon-1T excess. Bottom panels: Constraints from the Planck experiment on the DM annihilation to two $e^+e^-$ pair.}
	\label{fig3}
\end{figure}

Those constraints are summarized in Fig.~\ref{fig3}. Here CRESST-III (2019)~\cite{Abdelhameed:2019hmk} gives the most sensitive limits on the DM-nucleon scattering for the light mass range, 0.1$-$10 GeV. The DM-nucleon scattering cross-section against a nucleus with mass number $A$ and proton number $Z$ can be written as
\begin{equation}
\sigma_N=\frac{g_X^2 (Q_1+Q_2)^2 g_2^2 v_f^{\prime 2}}{16\pi\cos^2\theta}\frac{\mu^2_{DN}}{m^4_{\gamma'}}\left(\frac{Z}{A}\right)^2,
\end{equation}
where $\mu_{DN}$ is the DM-nucleon reduced mass and $Z/A\sim 0.5$. 

Dark photon experiments~\cite{Essig:2013lka} such as CHARM set limits on visible and invisible decays of the dark photon. These limits are shown in the kinetic mixing-dark photon mass plane in the top panels of Fig.~\ref{fig3} for benchmarks (a) (left), (b) (center) and (c) (right) recast to our model parameters. Also shown are the SuperCDMS~\cite{Agnese:2016cpb} sensitivity projections for benchmarks (a) and (b) while (c) remains out of reach of SuperCDMS. In the bottom panels, we present the recast limits from the Planck experiment~\cite{Ade:2015xua,Slatyer:2015jla} for the same benchmarks in the gauge coupling $g_X$-dark photon mass plane. These limits pertain to the annihilation of DM to two new vector bosons followed by their decay to two $e^+e^-$ pairs. 
The benchmarks of Table~\ref{tab1} satisfy all of the above constraints. There is a larger parameter space than the benchmarks of Table~\ref{tab1} where the experimental limits can be evaded while satisfying the relic density and producing the correct fit to the Xenon-1T excess. We exhibit those regions in the top panels of Fig.~\ref{fig3}. The green patches show the parts of the parameter space giving the correct DM relic density for a specific choice of the dark coupling $g_X$ and the grey stripes indicate the regions producing the correct fit to the Xenon-1T excess.

\begin{figure}[H]
 \centering
   \includegraphics[width=0.75\textwidth]{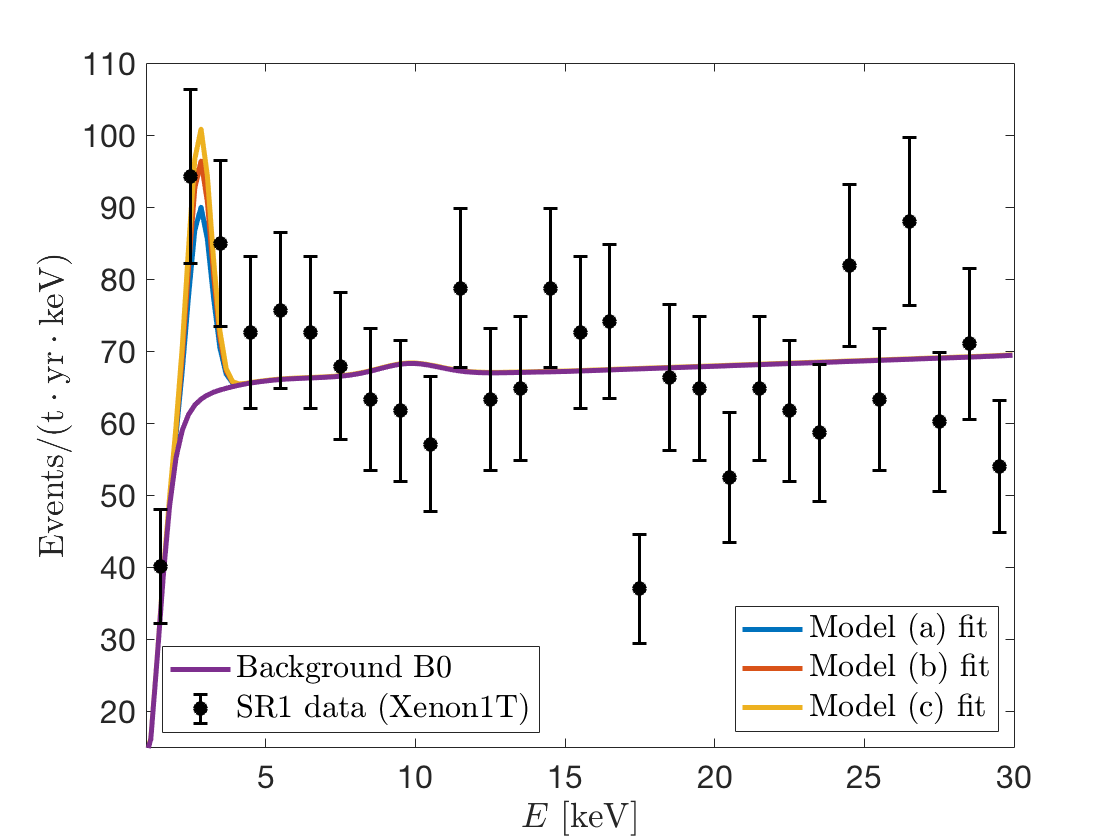}
   \caption{The event rate at Xenon-1T plotted over the background for benchmarks (purple line)
   vs. the electron recoil
   energy for models  (a) (blue line), (b) (red line)  and (c) (yellow line) of Table~\ref{tab1}.}
	\label{fig4}
\end{figure}

We exhibit in Fig.~\ref{fig4} the Xenon-1T data points (SR1) for the electron recoil events in the detector along with the background-only hypothesis (B0). The event rates for the three benchmarks in our model are plotted over the background showing a clear enhancement near 2.5$-$2.8 keV as expected. Lighter DM particles give a larger event rate since the latter is dependent on $\bar\sigma_e/m_2$. 
Thus more data in the future which can measure the height of the peak more accurately can
determine more precisely the allowed range of the dark matter mass.

\section{Conclusion\label{sec:conclu}}
In this work we have investigated the Xenon-1T signal as arising from  sub-GeV hidden sector dark 
matter. Specifically we consider a $U(1)$ Stueckelberg extension of the Standard Model with two
dark Dirac fermions degenerate in mass charged under the $U(1)$ gauge group. The fermion mass 
degeneracy is broken by a small mass mixing term which removes the degeneracy and produces 
two Dirac fermions $D_1', D_2'$ where $D_2'$ is heavier than $D_1'$ 
with a mass splitting 
size $\sim$ 3 keV. The observed effect is explained by the exothermic inelastic scattering
$eD_2'\to eD_1'$. The scattering occurs via exchange of a dark photon which is the massive 
gauge boson of the hidden sector. The coupling of the dark photon to the electron arises
from gauge kinetic mixing of the $U(1)$ gauge boson of the hidden sector with the gauge boson
of the $U(1)_Y$ hypercharge.
In the work here we have  given a detailed analysis of the dark matter relic density 
 constituted of dark fermions while the dark photons decay before the BBN time.
As noted above there are two  dark fermions in the system $D_1'$ and $D_2'$ where 
$D_2'$ has a mass slightly greater than that of $D_1'$ and decays to $D_1'$ via the 
channel $D'_2\to D_1'+ \nu\bar \nu$. However, the lifetime for this
decay is larger than the age of the universe and so for all practical 
purposes dark matter is constituted of  two dark fermions $D_2'$ and $D_1'$ in essentially equal amount.  In the analysis of the relic density one encounters  
 stringent constraints on kinetic mixing and on the dark photon mass  from 
 CRESST 2019 DM-nucleon scattering cross section, and  from the CHARM experiment. We have translated these constraints for our analysis for the model points considered in Table~\ref{tab1} 
 and show that the model parameters of Table~\ref{tab1} are consistent with these constraints (Fig.~\ref{fig3}).
 We note in passing that these constraints, i.e., on the kinetic mixing and on the dark photon mass,
 are projected to become more stringent 
 from future data from SuperCDMS as shown in Fig.~\ref{fig3} putting further constraints on the allowed
 parameter space of GeV size dark matter models. 
  Further,  the Planck experiment gives constraints on the 
 $U(1)_X$ gauge coupling and the dark photon mass from the dark matter annihilation to $4e$
 which arise in our case from the annihilation channel $D'\bar D'\to \gamma' \gamma'\to 4e$.
Here again we show that our model is consistent with these constraints as exhibited in Fig.~\ref{fig3}. 

In Fig.~\ref{fig4} we showed that the models listed in Table~\ref{tab1}, which satisfy all the known constraints 
on kinetic mixing, on the gauge coupling of the $U(1)_X$ of the dark sector, on the dark photon
mass, and generate relic density of dark matter consistent with the Planck experiment, 
can explain the Xenon-1T excess. We noted that the size of the peak for the excess events 
is model-dependent and more data in the future from the Xenon-1T collaboration will
help delineate the nature of the dark sector more accurately. Further checks on the model
can also come from additional data from the  direct detection experiments that focus
on the low mass region of dark matter in the GeV region.

\noindent
{\bf Acknowledgments:}  The research of AA and MK was supported by the BMBF under contract 05H18PMCC1, while the research of PN was supported in part by the NSF Grant PHY-1913328.

\section*{Appendix \label{sec:append}}
In this appendix we give further details of the analysis presented in the main body of the paper.
Thus in appendix A we discuss the generation of the mass term for the $D_1$ and $D_2$ quarks
and of the mixing term involving $\bar D_1 D_2+ \bar D_2 D_1$ from a Higgs mechanism.
In appendix B we give the cross-sections for the annihilation of $D\bar D$ into $q\bar q, \ell \bar \ell$,
$\nu\bar \nu$ and $\gamma' \gamma'$ and the cross section for the 
conversion process $D_2\bar D_2\to D_1\bar D_1$. In appendix C details of the partial decay
widths of the dark photon $\gamma'$ are given, i.e., $\gamma'\to \ell \bar\ell, q\bar q, \nu\bar \nu$. 

\appendix

\section{Generation of $\bar D_1D_2+ \bar D_2 D_1$ term from spontaneous symmetry breaking}
We now show that the term $\Delta \mu (\bar D_1 D_2+ \bar D_2D_1)$ can be  produced via spontaneous 
breaking which also gives mass to the $U(1)_X$ gauge boson. Thus consider a $U(1)_X$ gauge field
coupled to a complex scalar field charged under the $U(1)_X$ with charge $Q_\phi$, and further that
the complex scalar couples to the Dirac fermions $D_1, D_2$. Specifically we consider the 
Lagrangian 
\begin{align}
\mathcal{L}_{\phi}= -|(\partial_\mu \phi - ig_X Q_\phi C_\mu \phi)|^2 - V(\phi \phi^*)
- \lambda \frac{H^c H}{\Lambda} (\bar D_1 D_1 + \bar D_2D_2)
 -(\lambda'  \bar D_1 D_2 \phi + \text{h.c.}).
\end{align}
The Lagrangian above is invariant under $U(1)$ gauge transformations when 
\begin{align}
-Q_1+ Q_2 +Q_\phi=0,
\end{align}
where $Q_1$ and $Q_2$ are the $U(1)_X$ charges of $D_1$ and $D_2$. 
The potential $V(\phi)$ gives a VEV $\phi_0=\langle\phi\rangle$ and after spontaneous breaking 
the dark photon and the dark fermions get masses as follows
\begin{align}
\mathcal{L}_{\rm m}=  -\frac{1}{2} m^2_{\gamma'}  A^{\mu '}  A'_{\mu}
- \mu (\bar D_1 D_1 +\bar D_2 D_2)  -\Delta \mu 
(\bar D_1 D_2 + \bar D_2 D_1), 
\end{align} 
where 
\begin{align}
m_{\gamma'} &= \sqrt 2 g_X Q_\phi \phi_0, \nonumber\\
   \mu&= \lambda \frac{v^2}{\Lambda},\nonumber\\
\Delta \mu &= \lambda' \phi_0,
\end{align}
where $v\sim 250$ GeV is the standard model Higgs VEV. 
For $g_X\sim Q_\phi\sim 1$, an $m_{\gamma'}$ in the range 50$-$300 MeV requires 
$\phi_0\sim 100$ MeV. A Dirac fermion mass $\mu\sim 1$ GeV, requires the cutoff scale
$\Lambda\sim 100$ TeV.
The cutoff scale could have a low scale string origin. Further, $\Delta \mu \sim  2$ keV requires  
$\lambda'\sim 10^{-5}$. Such a small $\lambda'$ could also have a low scale string origin.
Thus quite remarkably if we assume that `flavor changing' $\Delta \mu$ term arises from a higher 
dimensional operator such as {$(H^cH/\Lambda^2) (\bar D_1 D_2\phi + \bar D_2 D_1\phi^*)$}, 
then after spontaneous breaking it produces a $\lambda'\sim v^2/\Lambda^2 \sim 10^{-5}$
which is precisely the size we want to generate $\Delta \mu\sim 2$ keV.

\section{Relevant cross-sections}

We present here the relevant cross-sections needed for the computation of the dark matter relic density. 
In the computations of these cross-sections, we have assumed $m_{D'_1}\approx m_{D'_2}\approx m_D$ as the mass difference between their masses is tiny and has no substantial effect on the 
size of the cross sections computed below.

\begin{enumerate}

\item $D\bar{D}\to Z/\gamma' \to q\bar{q}$: \\
The total cross-section for the process $D\bar{D}\to Z/\gamma' \to q\bar{q}$ is given by
\begin{align}
\sigma^{D\bar{D}\to q\bar{q}}(s)=&\frac{c^2_X g_X^2 g_2^2}{8\pi s\cos^2\theta}\sqrt{\frac{s-4m^2_q}{s-4m^2_D}}\Bigg[\frac{(\mathcal{R}_{12}-s_{\delta}\mathcal{R}_{22})^2(\alpha^2\eta_q T^2_{3q}-2\alpha\beta\kappa_q Q_q T_{3q}+2\beta^2 Q_q^2 \kappa_q)}{(s-m_Z^2)^2+m^2_Z\Gamma_Z^2} \nonumber \\
&+\frac{(\mathcal{R}_{11}-s_{\delta}\mathcal{R}_{21})^2(\alpha'^2\eta_q T^2_{3q}-2\alpha'\beta'\kappa_q  Q_q T_{3q}+2\beta'^2 Q_q^2 \kappa_q))}{(s-m^2_{\gamma'})^2+m^2_{\gamma'}\Gamma_{\gamma'}^2} \nonumber \\
&-2(\mathcal{R}_{11}-s_{\delta}\mathcal{R}_{21})(\mathcal{R}_{12}-s_{\delta}\mathcal{R}_{22})\Big\{Q_q\beta (2\beta' Q_q-\alpha' T_{3q})\kappa_q \nonumber \\
&+\alpha T_{3q}(\alpha' T_{3q}\eta_q-\beta' Q_q \kappa_q)\Big\} 
\times\frac{(s-m^2_Z)(s-m^2_{\gamma'})+\Gamma_Z\Gamma_{\gamma'}m_Z m_{\gamma'}}{[(s-m_Z^2)^2+m^2_Z\Gamma_Z^2][(s-m^2_{\gamma'})^2+m^2_{\gamma'}\Gamma_{\gamma'}^2]}\Bigg], 
\end{align}
where $m_q$, $m_Z$ and $m_{\gamma'}$ are the quark, $Z$ and $\gamma'$ masses, respectively, and $T_{3q}=1/2 (-1/2)$ and $Q_q=2/3 (-1/3)$ for up-(down)-type quarks, and with
\begin{equation}
\begin{aligned}
\kappa_q&=(s+2m^2_D)(s+2m^2_q),~~~\eta_q=(s+2m^2_D)(s-m^2_q), \\
\alpha&=\cos\psi-\bar{\epsilon}\sin\theta\sin\psi,~~~\beta=\sin^2\theta\cos\psi-\bar{\epsilon}\sin\theta\sin\psi, \\
\alpha'&=\sin\psi+\bar{\epsilon}\sin\theta\cos\psi,~~~\beta'=\sin^2\theta\sin\psi+\bar{\epsilon}\sin\theta\cos\psi.
\end{aligned}
\end{equation}

\item $D\bar{D}\to Z/\gamma' \to\ell\bar{\ell}$: \\
The total cross-section for the process $D\bar{D}\to Z/\gamma' \to\ell\bar{\ell}$ is given by
\begin{align}
\sigma^{D\bar{D}\to\ell\bar{\ell}}(s)=&\frac{c^2_X g_X^2 g_2^2}{96\pi s\cos^2\theta}\sqrt{\frac{s-4m^2_{\ell}}{s-4m^2_D}}\Bigg[\frac{(\mathcal{R}_{12}-s_{\delta}\mathcal{R}_{22})^2(\alpha^2\eta_{\ell}-4\alpha\beta\kappa_{\ell} +8\beta^2 \kappa_{\ell})}{(s-m_Z^2)^2+m^2_Z\Gamma_Z^2} \nonumber \\
&+\frac{(\mathcal{R}_{11}-s_{\delta}\mathcal{R}_{21})^2(\alpha'^2\eta_{\ell}-4\alpha'\beta'\kappa_{\ell} +8\beta'^2 \kappa_{\ell})}{(s-m_{\gamma'}^2)^2+m^2_{\gamma'}\Gamma_{\gamma'}^2} \nonumber \\
&+2(\mathcal{R}_{11}-s_{\delta}\mathcal{R}_{21})(\mathcal{R}_{12}-s_{\delta}\mathcal{R}_{22})\Big\{2\beta(\alpha'-4\beta')\kappa_{\ell} \nonumber \\
&-\alpha (\alpha'\eta_{\ell}-2\beta' \kappa_{\ell})\Big\}\times      \frac{(s-m^2_Z)(s-m^2_{Z'})+\Gamma_Z\Gamma_{\gamma'}m_Z m_{\gamma'}}{[(s-m_Z^2)^2+m^2_Z\Gamma_Z^2][(s-m^2_{\gamma'})^2+m^2_{\gamma'}\Gamma_{\gamma'}^2]} \Bigg], 
\end{align}
where
\begin{align}
\kappa_{\ell}&=(s+2m^2_D)(s+2m^2_{\ell}),~~~\eta_{\ell}=(s+2m^2_D)(s-m^2_{\ell}).
\end{align}

\item $D\bar{D}\to Z/\gamma' \to\nu\bar{\nu}$: \\
The total cross-section for the process $D\bar{D}\to Z/\gamma' \to\nu\bar{\nu}$ is given by
\begin{align}
\sigma^{D\bar{D}\to\nu\bar{\nu}}(s)&=\frac{c^2_X g_X^2 g^2_2 }{32\pi\cos^2\theta}\frac{(s+2m^2_D)s^{1/2}}{\sqrt{s-4m^2_D}}\Bigg\{\frac{\alpha'^2(\mathcal{R}_{11}-s_{\delta}\mathcal{R}_{21})^2}{(s-m_{\gamma'}^2)^2+m^2_{\gamma'}\Gamma_{\gamma'}^2}+\frac{\alpha^2(\mathcal{R}_{12}-s_{\delta}\mathcal{R}_{22})^2}{(s-m_{Z}^2)^2+m^2_{Z}\Gamma_{Z}^2} \nonumber \\
&-2\alpha\alpha'(\mathcal{R}_{11}-s_{\delta}\mathcal{R}_{21})(\mathcal{R}_{12}-s_{\delta}\mathcal{R}_{22}) \nonumber \\
&\times\frac{(s-m^2_Z)(s-m^2_{\gamma'})+\Gamma_Z\Gamma_{\gamma'}m_Z m_{\gamma'}}{[(s-m_Z^2)^2+m^2_Z\Gamma_Z^2][(s-m^2_{\gamma'})^2+m^2_{\gamma'}\Gamma_{\gamma'}^2]} \Bigg\}.
\end{align}
In all of the above, the coefficient $c_X$ is defined as
\begin{equation}
c_X= 
\begin{cases}
    \frac{1}{2}(Q_1+Q_2),& \text{for } \bar{D}'_1 D'_1/\bar{D}'_2 D'_2\\
    \frac{1}{2}(Q_1-Q_2), & \text{for } \bar{D}'_1 D'_2/\bar{D}'_2 D'_1.
\end{cases}
\end{equation}

\item $D'_{i}\bar{D}'_{j}\longleftrightarrow \gamma' \gamma'$: \\
The total cross-section for the process $D'_{i}\bar{D}'_{j}\to \gamma' \gamma'$ is
\begin{equation}
\sigma^{D'_{i}\bar{D}'_{j}\to \gamma' \gamma'}(s)=\sum_{ij}c_{ij}^4\sigma_0(s),
\end{equation}
where
\begin{align}
\sigma_0(s)=\frac{g_{X}^4(\mathcal{R}_{11}-s_{\delta}\mathcal{R}_{21})^4}{8\pi s(s-4m^2_D)}&\Bigg\{-\frac{\sqrt{(s-4m^2_{\gamma'})(s-4m^2_{D})}}{m^4_{\gamma'}+m^2_D(s-4m^2_{\gamma'})}[2m^4_{\gamma'}+m^2_D(s+4m^2_D)] \nonumber \\
&+\frac{\log A}{s-2m^2_{\gamma'}}(s^2+4m^2_D s+4m^4_{\gamma'}-8m^4_D-8m_D^2 m^2_{\gamma'})\Bigg\},
\end{align}
and
\begin{equation}
A=\frac{s-2m^2_{\gamma'}+\sqrt{(s-4m^2_{\gamma'})(s-4m^2_{D})}}{s-2m^2_{\gamma'}-\sqrt{(s-4m^2_{\gamma'})(s-4m^2_{D})}},
\end{equation}
with
\begin{equation}
 c_{ij}= 
\begin{cases}
   \frac{1}{2}(Q_1+Q_2),& \text{for } i=j=1,2 \\
    \frac{1}{2}\sqrt{(Q_1+Q_2)(Q_1-Q_2)}, & \text{for } i,j=\{1,2\}, i\neq j.
\end{cases}
\end{equation}

The reverse processes are given by
\begin{equation}
9(s-4m^2_{\gamma'})\sigma^{\gamma' \gamma'\to D'_{i}\bar{D}'_{j}}(s)=8(s-4m^2_{D})\sigma^{D'_{i}\bar{D}'_{j}\to \gamma'\gamma'}(s).
\end{equation}

\item The conversion process $D_2'\bar{D_2'}\longrightarrow D_1'\bar{D_1'}$: \\
\begin{equation}
\sigma^{D_2'\bar{D_2'}\longrightarrow D_1'\bar{D_1'}}\simeq \frac{g_X^4(Q_1+Q_2)^4}{4\pi}\frac{m^2_D}{m^4_{\gamma'}}\frac{(1-5r+7r^2)}{(1-r)^2},
\end{equation}
where $r=m^2_{\gamma'}/4m^2_D$.

\end{enumerate}
 
\section{Decay widths for the processes $\gamma'\to\ell\bar{\ell},q\bar{q},\nu\bar{\nu}$}

\begin{enumerate}

\item The decay width of $\gamma'$ to leptons is given by
\begin{align}
\Gamma_{\gamma'\to\ell\bar\ell}=\frac{g_2^2}{24\pi m_{\gamma'}\cos^2\theta}\sqrt{1-\left(\frac{2m_{\ell}}{m_{\gamma'}}\right)^2}&\Bigg[\frac{1}{4}\alpha'^2(m^2_{\gamma'}-m^2_{\ell})-\alpha'\beta'(m^2_{\gamma'}+2m^2_{\ell})\nonumber \\
&+2\beta'^2(m^2_{\gamma'}+2m^2_{\ell})\Bigg].
\end{align}

\item The decay width of $\gamma'$ to quarks is given by
\begin{align}
\Gamma_{\gamma'\to q\bar q}=\frac{g_2^2}{8\pi m_{\gamma'}\cos^2\theta}\sqrt{1-\left(\frac{2m_{q}}{m_{\gamma'}}\right)^2}&[\alpha'^2(m^2_{\gamma'}-m^2_{q})T_{3q}^2-2\alpha'\beta'(m^2_{\gamma'}+2m^2_{q})Q_q T_{3q}\nonumber \\
&+2\beta'^2(m^2_{\gamma'}+2m^2_{q})Q_q^2].
\end{align}

\item The decay width of $\gamma'$ to neutrinos is given by
\begin{equation}
\Gamma_{\gamma'\to\nu\bar\nu}=\frac{g_2^2}{32\pi\cos^2\theta}m_{\gamma'}\alpha'^2.
\end{equation}

\end{enumerate}


\begin{thebibliography}{999}

%\cite{Aprile:2020tmw}
\bibitem{Aprile:2020tmw}
E.~Aprile \textit{et al.} [XENON],
%``Excess electronic recoil events in XENON1T,''
Phys. Rev. D \textbf{102}, no.7, 072004 (2020)
doi:10.1103/PhysRevD.102.072004
[arXiv:2006.09721 [hep-ex]].
%155 citations counted in INSPIRE as of 16 Nov 2020

%\cite{Viaux:2013lha}
\bibitem{Viaux:2013lha}
N.~Viaux, M.~Catelan, P.~B.~Stetson, G.~Raffelt, J.~Redondo, A.~A.~R.~Valcarce and A.~Weiss,
%``Neutrino and axion bounds from the globular cluster M5 (NGC 5904),''
Phys. Rev. Lett. \textbf{111}, 231301 (2013)
doi:10.1103/PhysRevLett.111.231301
[arXiv:1311.1669 [astro-ph.SR]].
%158 citations counted in INSPIRE as of 04 Nov 2020

%\cite{Bertolami:2014wua}
\bibitem{Bertolami:2014wua}
M.~M.~Miller Bertolami, B.~E.~Melendez, L.~G.~Althaus and J.~Isern,
%``Revisiting the axion bounds from the Galactic white dwarf luminosity function,''
JCAP \textbf{10}, 069 (2014)
doi:10.1088/1475-7516/2014/10/069
[arXiv:1406.7712 [hep-ph]].
%109 citations counted in INSPIRE as of 26 Oct 2020

%\cite{Battich:2016htm}
\bibitem{Battich:2016htm}
T.~Battich, A.~H.~C\'orsico, L.~G.~Althaus, M.~M.~Miller Bertolami and M.~M.~M.~Bertolami,
%``First axion bounds from a pulsating helium-rich white dwarf star,''
JCAP \textbf{08}, 062 (2016)
doi:10.1088/1475-7516/2016/08/062
[arXiv:1605.07668 [astro-ph.SR]].
%27 citations counted in INSPIRE as of 26 Oct 2020

%\cite{Giannotti:2017hny}
\bibitem{Giannotti:2017hny}
M.~Giannotti, I.~G.~Irastorza, J.~Redondo, A.~Ringwald and K.~Saikawa,
%``Stellar Recipes for Axion Hunters,''
JCAP \textbf{10}, 010 (2017)
doi:10.1088/1475-7516/2017/10/010
[arXiv:1708.02111 [hep-ph]].
%121 citations counted in INSPIRE as of 04 Nov 2020

%\cite{Shakeri:2020wvk}
\bibitem{Shakeri:2020wvk}
S.~Shakeri, F.~Hajkarim and S.~S.~Xue,
%``Shedding New Light on Sterile Neutrinos from XENON1T Experiment,''
[arXiv:2008.05029 [hep-ph]].
%4 citations counted in INSPIRE as of 05 Nov 2020

%\cite{Khruschov:2020cnf}
\bibitem{Khruschov:2020cnf}
V.~V.~Khruschov,
%``Interpretation of the XENON1T excess in the model with decaying sterile neutrinos,''
[arXiv:2008.03150 [hep-ph]].
%2 citations counted in INSPIRE as of 15 Oct 2020

%\cite{Arcadi:2020zni}
\bibitem{Arcadi:2020zni}
G.~Arcadi, A.~Bally, F.~Goertz, K.~Tame-Narvaez, V.~Tenorth and S.~Vogl,
%``EFT Interpretation of XENON1T Electron Recoil Excess: Neutrinos and Dark Matter,''
[arXiv:2007.08500 [hep-ph]].
%11 citations counted in INSPIRE as of 15 Oct 2020

%\cite{Khan:2020vaf}
\bibitem{Khan:2020vaf}
A.~N.~Khan,
%``Can Nonstandard Neutrino Interactions explain the XENON1T spectral excess?,''
Phys. Lett. B \textbf{809}, 135782 (2020)
doi:10.1016/j.physletb.2020.135782
[arXiv:2006.12887 [hep-ph]].
%41 citations counted in INSPIRE as of 06 Nov 2020

%\cite{Cao:2020oxq}
\bibitem{Cao:2020oxq}
J.~Cao, X.~Du, Z.~Li, F.~Wang and Y.~Zhang,
%``Explaining The XENON1T Excess With Light Goldstini Dark Matter,''
[arXiv:2007.09981 [hep-ph]].
%5 citations counted in INSPIRE as of 15 Oct 2020

%\cite{Takahashi:2020uio}
\bibitem{Takahashi:2020uio}
F.~Takahashi, M.~Yamada and W.~Yin,
%``What if ALP dark matter for the XENON1T excess is the inflaton,''
[arXiv:2007.10311 [hep-ph]].
%6 citations counted in INSPIRE as of 15 Oct 2020

%\cite{Karozas:2020pun}
\bibitem{Karozas:2020pun}
A.~Karozas, S.~F.~King, G.~K.~Leontaris and D.~K.~Papoulias,
%``Low Scale String Theory Benchmarks for Hidden Photon Dark Matter Interpretations of the XENON1T Anomaly,''
[arXiv:2008.03295 [hep-ph]].
%1 citations counted in INSPIRE as of 15 Oct 2020

%\cite{Anchordoqui:2020tlp}
\bibitem{Anchordoqui:2020tlp}
L.~A.~Anchordoqui, I.~Antoniadis, K.~Benakli and D.~Lust,
%``Anomalous $U(1)$ gauge bosons as light dark matter in string theory,''
Phys. Lett. B \textbf{810}, 135838 (2020)
doi:10.1016/j.physletb.2020.135838
[arXiv:2007.11697 [hep-th]].
%5 citations counted in INSPIRE as of 18 Nov 2020


%\cite{Jho:2020sku}
\bibitem{Jho:2020sku}
Y.~Jho, J.~C.~Park, S.~C.~Park and P.~Y.~Tseng,
%``Leptonic New Force and Cosmic-ray Boosted Dark Matter for the XENON1T Excess,''
Phys. Lett. B \textbf{811}, 135863 (2020)
doi:10.1016/j.physletb.2020.135863
[arXiv:2006.13910 [hep-ph]].
%36 citations counted in INSPIRE as of 22 Oct 2020

%\cite{Fornal:2020npv}
\bibitem{Fornal:2020npv}
B.~Fornal, P.~Sandick, J.~Shu, M.~Su and Y.~Zhao,
%``Boosted Dark Matter Interpretation of the XENON1T Excess,''
Phys. Rev. Lett. \textbf{125}, no.16, 161804 (2020)
doi:10.1103/PhysRevLett.125.161804
[arXiv:2006.11264 [hep-ph]].
%63 citations counted in INSPIRE as of 06 Nov 2020

%\cite{DelleRose:2020pbh}
\bibitem{DelleRose:2020pbh}
L.~Delle Rose, G.~H\"utsi, C.~Marzo and L.~Marzola,
%``Impact of loop-induced processes on the boosted dark matter interpretation of the XENON1T excess,''
[arXiv:2006.16078 [hep-ph]].
%19 citations counted in INSPIRE as of 18 Nov 2020

%\cite{Millea:2020xxp}
\bibitem{Millea:2020xxp}
M.~Millea,
%``New cosmological bounds on axions in the XENON1T window,''
[arXiv:2007.05659 [astro-ph.CO]].
%3 citations counted in INSPIRE as of 31 Oct 2020

%\cite{Arias-Aragon:2020qtn}
\bibitem{Arias-Aragon:2020qtn}
F.~Arias-Aragon, F.~D'Eramo, R.~Z.~Ferreira, L.~Merlo and A.~Notari,
%``Cosmic Imprints of XENON1T Axions,''
[arXiv:2007.06579 [hep-ph]].
%8 citations counted in INSPIRE as of 15 Oct 2020

%\cite{Long:2020uyf}
\bibitem{Long:2020uyf}
H.~N.~Long, D.~V.~Soa, V.~H.~Binh and A.~E.~C\'arcamo Hern\'andez,
%``Linking axion-like dark matter, the XENON1T excess, inflation and the tiny active neutrino masses,''
[arXiv:2007.05004 [hep-ph]].
%4 citations counted in INSPIRE as of 04 Nov 2020

%\cite{Athron:2020maw}
\bibitem{Athron:2020maw}
P.~Athron, C.~Bal\'azs, A.~Beniwal, J.~E.~Camargo-Molina, A.~Fowlie, T.~E.~Gonzalo, S.~Hoof, F.~Kahlhoefer, D.~J.~E.~Marsh and M.~T.~Prim, \textit{et al.}
%``Global fits of axion-like particles to XENON1T and astrophysical data,''
[arXiv:2007.05517 [astro-ph.CO]].
%17 citations counted in INSPIRE as of 06 Nov 2020

%\cite{Li:2020naa}
\bibitem{Li:2020naa}
T.~Li,
%``The KSVZ Axion and Pseudo-Nambu-Goldstone Boson Models for the XENON1T Excess,''
[arXiv:2007.00874 [hep-ph]].
%17 citations counted in INSPIRE as of 15 Oct 2020

%\cite{Cacciapaglia:2020kbf}
\bibitem{Cacciapaglia:2020kbf}
C.~Cai, H.~H.~Zhang, M.~T.~Frandsen, M.~Rosenlyst and G.~Cacciapaglia,
%``XENON1T solar axion and the Higgs boson emerging from the dark,''
Phys. Rev. D \textbf{102}, no.7, 075018 (2020)
doi:10.1103/PhysRevD.102.075018
[arXiv:2006.16267 [hep-ph]].
%18 citations counted in INSPIRE as of 06 Nov 2020

%\cite{Gao:2020wer}
\bibitem{Gao:2020wer}
C.~Gao, J.~Liu, L.~T.~Wang, X.~P.~Wang, W.~Xue and Y.~M.~Zhong,
%``Reexamining the Solar Axion Explanation for the XENON1T Excess,''
Phys. Rev. Lett. \textbf{125}, no.13, 131806 (2020)
doi:10.1103/PhysRevLett.125.131806
[arXiv:2006.14598 [hep-ph]].
%43 citations counted in INSPIRE as of 16 Nov 2020

%\cite{Chiang:2020hgb}
\bibitem{Chiang:2020hgb}
C.~W.~Chiang and B.~Q.~Lu,
%``Evidence of A Simple Dark Sector from XENON1T Anomaly,''
[arXiv:2007.06401 [hep-ph]].
%7 citations counted in INSPIRE as of 15 Oct 2020

%\cite{Choi:2020kch}
\bibitem{Choi:2020kch}
G.~Choi, T.~T.~Yanagida and N.~Yokozaki,
%``Feebly interacting $U (1)_{B−L}$ gauge boson warm dark matter and XENON1T anomaly,''
Phys. Lett. B \textbf{810}, 135836 (2020)
doi:10.1016/j.physletb.2020.135836
[arXiv:2007.04278 [hep-ph]].
%10 citations counted in INSPIRE as of 26 Oct 2020

%\cite{Okada:2020evk}
\bibitem{Okada:2020evk}
N.~Okada, S.~Okada, D.~Raut and Q.~Shafi,
%``Dark matter $Z^\prime$ and XENON1T excess from $U(1)_X$ extended standard model,''
Phys. Lett. B \textbf{810}, 135785 (2020)
doi:10.1016/j.physletb.2020.135785
[arXiv:2007.02898 [hep-ph]].
%21 citations counted in INSPIRE as of 09 Nov 2020

%\cite{Baek:2020owl}
\bibitem{Baek:2020owl}
S.~Baek, J.~Kim and P.~Ko,
%``XENON1T excess in local $Z_2$ DM models with light dark sector,''
Phys. Lett. B \textbf{810}, 135848 (2020)
doi:10.1016/j.physletb.2020.135848
[arXiv:2006.16876 [hep-ph]].
%21 citations counted in INSPIRE as of 30 Oct 2020

%\cite{Gao:2020wfr}
\bibitem{Gao:2020wfr}
Y.~Gao and T.~Li,
%``Lepton Number Violating Electron Recoils at XENON1T by the $U(1)_{B-L}$ Model with Non-Standard Interactions,''
[arXiv:2006.16192 [hep-ph]].
%19 citations counted in INSPIRE as of 01 Nov 2020

%\cite{Lindner:2020kko}
\bibitem{Lindner:2020kko}
M.~Lindner, Y.~Mambrini, T.~B.~d.~Melo and F.~S.~Queiroz,
%``XENON1T Anomaly: A Light $Z^\prime$,''
[arXiv:2006.14590 [hep-ph]].
%38 citations counted in INSPIRE as of 16 Nov 2020

%\cite{Bramante:2020zos}
\bibitem{Bramante:2020zos}
J.~Bramante and N.~Song,
%``Electric But Not Eclectic: Thermal Relic Dark Matter for the XENON1T Excess,''
Phys. Rev. Lett. \textbf{125}, no.16, 161805 (2020)
doi:10.1103/PhysRevLett.125.161805
[arXiv:2006.14089 [hep-ph]].
%38 citations counted in INSPIRE as of 22 Oct 2020

%\cite{AristizabalSierra:2020edu}
\bibitem{AristizabalSierra:2020edu}
D.~Aristizabal Sierra, V.~De Romeri, L.~J.~Flores and D.~K.~Papoulias,
%``Light vector mediators facing XENON1T data,''
Phys. Lett. B \textbf{809}, 135681 (2020)
doi:10.1016/j.physletb.2020.135681
[arXiv:2006.12457 [hep-ph]].
%52 citations counted in INSPIRE as of 16 Nov 2020

%\cite{Harigaya:2020ckz}
\bibitem{Harigaya:2020ckz}
K.~Harigaya, Y.~Nakai and M.~Suzuki,
%``Inelastic Dark Matter Electron Scattering and the XENON1T Excess,''
Phys. Lett. B \textbf{809}, 135729 (2020)
doi:10.1016/j.physletb.2020.135729
[arXiv:2006.11938 [hep-ph]].
%45 citations counted in INSPIRE as of 15 Oct 2020

%\cite{Bell:2020bes}
\bibitem{Bell:2020bes}
N.~F.~Bell, J.~B.~Dent, B.~Dutta, S.~Ghosh, J.~Kumar and J.~L.~Newstead,
%``Explaining the XENON1T excess with Luminous Dark Matter,''
Phys. Rev. Lett. \textbf{125}, no.16, 161803 (2020)
doi:10.1103/PhysRevLett.125.161803
[arXiv:2006.12461 [hep-ph]].
%52 citations counted in INSPIRE as of 06 Nov 2020

%\cite{Du:2020ybt}
\bibitem{Du:2020ybt}
M.~Du, J.~Liang, Z.~Liu, V.~Tran and Y.~Xue,
%``On-shell mediator dark matter models and the Xenon1T anomaly,''
[arXiv:2006.11949 [hep-ph]].
%40 citations counted in INSPIRE as of 15 Oct 2020

%\cite{Chakraborty:2020vec}
\bibitem{Chakraborty:2020vec}
S.~Chakraborty, T.~H.~Jung, V.~Loladze, T.~Okui and K.~Tobioka,
%``A solar origin of the XENON1T excess without stellar cooling problems,''
[arXiv:2008.10610 [hep-ph]].
%1 citations counted in INSPIRE as of 15 Oct 2020

%\cite{Borah:2020jzi}
\bibitem{Borah:2020jzi}
D.~Borah, S.~Mahapatra, D.~Nanda and N.~Sahu,
%``Inelastic Fermion Dark Matter Origin of XENON1T Excess with Muon $(g-2)$ and Light Neutrino Mass,''
[arXiv:2007.10754 [hep-ph]].
%7 citations counted in INSPIRE as of 07 Nov 2020

%\cite{Bally:2020yid}
\bibitem{Bally:2020yid}
A.~Bally, S.~Jana and A.~Trautner,
%``Neutrino self-interactions and XENON1T electron recoil excess,''
Phys. Rev. Lett. \textbf{125}, no.16, 161802 (2020)
doi:10.1103/PhysRevLett.125.161802
[arXiv:2006.11919 [hep-ph]].
%51 citations counted in INSPIRE as of 18 Nov 2020

%\cite{Kim:2020aua}
\bibitem{Kim:2020aua}
J.~Kim, T.~Nomura and H.~Okada,
%``A radiative seesaw model linking to XENON1T anomaly,''
Phys. Lett. B \textbf{811}, 135862 (2020)
doi:10.1016/j.physletb.2020.135862
[arXiv:2007.09894 [hep-ph]].
%6 citations counted in INSPIRE as of 18 Nov 2020

%\cite{Farzan:2020dds}
\bibitem{Farzan:2020dds}
Y.~Farzan and M.~Rajaee,
%``Pico-charged particles explaining 511 keV line and XENON1T signal,''
[arXiv:2007.14421 [hep-ph]].
%2 citations counted in INSPIRE as of 18 Nov 2020

%\cite{Choudhury:2020xui}
\bibitem{Choudhury:2020xui}
D.~Choudhury, S.~Maharana, D.~Sachdeva and V.~Sahdev,
%``Dark Matter, Muon Anomalous Magnetic Moment and the XENON1T Excess,''
Phys. Rev. D \textbf{103}, no.1, 015006 (2021)
doi:10.1103/PhysRevD.103.015006
[arXiv:2007.08205 [hep-ph]].
%11 citations counted in INSPIRE as of 18 Jan 2021

%\cite{Babu:2020ivd}
\bibitem{Babu:2020ivd}
K.~S.~Babu, S.~Jana and M.~Lindner,
%``Large Neutrino Magnetic Moments in the Light of Recent Experiments,''
JHEP \textbf{10}, 040 (2020)
doi:10.1007/JHEP10(2020)040
[arXiv:2007.04291 [hep-ph]].
%14 citations counted in INSPIRE as of 19 Jan 2021

%\cite{Bergsma:1985is}
\bibitem{Bergsma:1985is}
F.~Bergsma \textit{et al.} [CHARM],
%``A Search for Decays of Heavy Neutrinos in the Mass Range 0.5-\{GeV\} to 2.8-\{GeV\},''
Phys. Lett. B \textbf{166}, 473-478 (1986)
doi:10.1016/0370-2693(86)91601-1
%227 citations counted in INSPIRE as of 11 Nov 2020

%\cite{Gninenko:2012eq}
\bibitem{Gninenko:2012eq}
S.~N.~Gninenko,
%``Constraints on sub-GeV hidden sector gauge bosons from a search for heavy neutrino decays,''
Phys. Lett. B \textbf{713}, 244-248 (2012)
doi:10.1016/j.physletb.2012.06.002
[arXiv:1204.3583 [hep-ph]].
%102 citations counted in INSPIRE as of 05 Nov 2020

%\cite{Holdom:1985ag}
\bibitem{Holdom:1985ag}
B.~Holdom,
%``Two U(1)'s and Epsilon Charge Shifts,''
Phys. Lett. B \textbf{166}, 196-198 (1986)
doi:10.1016/0370-2693(86)91377-8;
  %%CITATION = doi:10.1016/0370-2693(86)91377-8;%%
  %1599 citations counted in INSPIRE as of 23 Feb 2020;
%\bibitem{Holdom:1991}
 % B. Holdom,
  Phys.\ Lett.\ B {\bf 259}, 329 (1991).
  doi:10.1016/0370-2693(91)90836-F
  %%CITATION = doi:10.1016/0370-2693(91)90836-F;%%
  %165 citations counted in INSPIRE as of 18 May 2018

%\cite{Kors:2004dx}
\bibitem{Kors:2004dx}
B.~Kors and P.~Nath,
%``A Stueckelberg extension of the standard model,''
Phys. Lett. B \textbf{586}, 366-372 (2004)
doi:10.1016/j.physletb.2004.02.051
[arXiv:hep-ph/0402047 [hep-ph]];
%196 citations counted in INSPIRE as of 15 Oct 2020
JHEP \textbf{07}, 069 (2005)
doi:10.1088/1126-6708/2005/07/069.
%[arXiv:hep-ph/0503208 [hep-ph]].
%149 citations counted in INSPIRE as of 11 Nov 2020


%\cite{Cheung:2007ut}
\bibitem{Cheung:2007ut}
K.~Cheung and T.~C.~Yuan,
%``Hidden fermion as milli-charged dark matter in Stueckelberg Z- prime model,''
JHEP \textbf{03}, 120 (2007)
doi:10.1088/1126-6708/2007/03/120
[arXiv:hep-ph/0701107 [hep-ph]].
%91 citations counted in INSPIRE as of 25 Oct 2020

%\cite{Feldman:2007wj}
\bibitem{Feldman:2007wj}
D.~Feldman, Z.~Liu and P.~Nath,
%``The Stueckelberg Z-prime Extension with Kinetic Mixing and Milli-Charged Dark Matter From the Hidden Sector,''
Phys. Rev. D \textbf{75}, 115001 (2007)
doi:10.1103/PhysRevD.75.115001
[arXiv:hep-ph/0702123 [hep-ph]];
%286 citations counted in INSPIRE as of 09 Nov 2020
%D.~Feldman, Z.~Liu and P.~Nath,
%``PAMELA Positron Excess as a Signal from the Hidden Sector,''
Phys. Rev. D \textbf{79}, 063509 (2009)
doi:10.1103  /PhysRevD.79.063509.
%[arXiv:0810.5762 [hep-ph]].
%217 citations counted in INSPIRE as of 12 Nov 2020



%\cite{Hall:2009bx}
\bibitem{Hall:2009bx}
L.~J.~Hall, K.~Jedamzik, J.~March-Russell and S.~M.~West,
%``Freeze-In Production of FIMP Dark Matter,''
JHEP \textbf{03}, 080 (2010)
doi:10.1007/JHEP03(2010)080
[arXiv:0911.1120 [hep-ph]].
%635 citations counted in INSPIRE as of 16 Nov 2020

%\cite{Aboubrahim:2020lnr}
\bibitem{Aboubrahim:2020lnr}
A.~Aboubrahim, W.~Z.~Feng, P.~Nath and Z.~Y.~Wang,
%``Self-interacting hidden sector dark matter and small scale galaxy structure anomalies,''
[arXiv:2008.00529 [hep-ph]].
%0 citations counted in INSPIRE as of 15 Oct 2020

%\cite{Aghanim:2018eyx}
\bibitem{Aghanim:2018eyx}
N.~Aghanim \textit{et al.} [Planck],
%``Planck 2018 results. VI. Cosmological parameters,''
Astron. Astrophys. \textbf{641}, A6 (2020)
doi:10.1051/0004-6361/201833910
[arXiv:1807.06209 [astro-ph.CO]].
%3869 citations counted in INSPIRE as of 16 Nov 2020

%\cite{Lee:2020wmh}
\bibitem{Lee:2020wmh}
H.~M.~Lee,
%``Exothermic Dark Matter for XENON1T Excess,''
[arXiv:2006.13183 [hep-ph]].
%36 citations counted in INSPIRE as of 15 Oct 2020

%\cite{Roberts:2019chv}
\bibitem{Roberts:2019chv}
B.~M.~Roberts and V.~V.~Flambaum,
%``Electron-interacting dark matter: Implications from DAMA/LIBRA-phase2 and prospects for liquid xenon detectors and NaI detectors,''
Phys. Rev. D \textbf{100}, no.6, 063017 (2019)
doi:10.1103/PhysRevD.100.063017
[arXiv:1904.07127 [hep-ph]].
%17 citations counted in INSPIRE as of 15 Oct 2020

%\cite{Abdelhameed:2019hmk}
\bibitem{Abdelhameed:2019hmk}
A.~H.~Abdelhameed \textit{et al.} [CRESST],
%``First results from the CRESST-III low-mass dark matter program,''
Phys. Rev. D \textbf{100}, no.10, 102002 (2019)
doi:10.1103/PhysRevD.100.102002
[arXiv:1904.00498 [astro-ph.CO]].
%108 citations counted in INSPIRE as of 06 Nov 2020

%\cite{Essig:2013lka}
\bibitem{Essig:2013lka}
R.~Essig, J.~A.~Jaros, W.~Wester, P.~Hansson Adrian, S.~Andreas, T.~Averett, O.~Baker, B.~Batell, M.~Battaglieri and J.~Beacham, \textit{et al.}
%``Working Group Report: New Light Weakly Coupled Particles,''
[arXiv:1311.0029 [hep-ph]].
%539 citations counted in INSPIRE as of 16 Nov 2020

%\cite{Agnese:2016cpb}
\bibitem{Agnese:2016cpb}
R.~Agnese \textit{et al.} [SuperCDMS],
%``Projected Sensitivity of the SuperCDMS SNOLAB experiment,''
Phys. Rev. D \textbf{95}, no.8, 082002 (2017)
doi:10.1103/PhysRevD.95.082002
[arXiv:1610.00006 [physics.ins-det]].
%201 citations counted in INSPIRE as of 05 Nov 2020

%\cite{Ade:2015xua}
\bibitem{Ade:2015xua}
P.~A.~R.~Ade \textit{et al.} [Planck],
%``Planck 2015 results. XIII. Cosmological parameters,''
Astron. Astrophys. \textbf{594}, A13 (2016)
doi:10.1051/0004-6361/201525830
[arXiv:1502.01589 [astro-ph.CO]].
%9303 citations counted in INSPIRE as of 16 Nov 2020

%\cite{Slatyer:2015jla}
\bibitem{Slatyer:2015jla}
T.~R.~Slatyer,
%``Indirect dark matter signatures in the cosmic dark ages. I. Generalizing the bound on s-wave dark matter annihilation from Planck results,''
Phys. Rev. D \textbf{93}, no.2, 023527 (2016)
doi:10.1103/PhysRevD.93.023527
[arXiv:1506.03811 [hep-ph]].
%253 citations counted in INSPIRE as of 11 Nov 2020

\end{thebibliography}
\end{document}